\documentclass[sigconf]{acmart}
\acmConference[ICPC 2022]{The 30th International Conference on Program Comprehension}{May 21–22, 2022}{Pittsburgh, PA, USA}
\usepackage{graphicx}
\usepackage{subfigure}
\usepackage{multirow}
\usepackage{multicol}
\usepackage[breakable]{tcolorbox}
\usepackage{pifont}
\usepackage{enumitem}
\usepackage{amsmath}

\usepackage[normalem]{ulem}
\useunder{\uline}{\ul}{}
\usepackage{threeparttable}
\usepackage{makecell}

\usepackage{xspace}
\makeatletter
\DeclareRobustCommand\onedot{\futurelet\@let@token\@onedot}
\def\@onedot{\ifx\@let@token.\else.\null\fi\xspace}
\def\eg{\emph{e.g}\onedot} \def\Eg{\emph{E.g}\onedot}
\def\ie{\emph{i.e}\onedot} 
 
\def\etc{\emph{etc}\onedot}

\makeatother

\renewcommand{\paragraph}[1]{\vskip 0.01in \noindent {\bf #1.}}

\newcommand{\maydelete}[1]{\textcolor{red}{maydelete: #1}}

\newcommand{\wwh}[1]{\textcolor{orange}{wwh: #1}}

\newcommand{\citea}[1]{\citeauthor{#1}~\cite{#1}}

\AtBeginDocument{%
  \providecommand\BibTeX{{%
    \normalfont B\kern-0.5em{\scshape i\kern-0.25em b}\kern-0.8em\TeX}}}

\setcopyright{acmcopyright}
\copyrightyear{2022}
\acmYear{2022}
\setcopyright{acmcopyright}\acmConference[ICPC '22]{30th International Conference on Program Comprehension}{May 16--17, 2022}{Virtual Event, USA}
\acmBooktitle{30th International Conference on Program Comprehension (ICPC '22), May 16--17, 2022, Virtual Event, USA}
\acmPrice{15.00}
\acmDOI{10.1145/3524610.3527905}
\acmISBN{978-1-4503-9298-3/22/05}

\begin{document}
%%
%% The "title" command has an optional parameter,
%% allowing the author to define a "short title" to be used in page headers.
% \title{Represent Programs with Heterogeneous Graphs}
% \title{Learning Representation for Programs by Exploring the Heterogeneity of Abstract Syntax Trees}
% \title{Learning Heterogeneous Type Information in Program Graphs}
\title{Learning to Represent Programs with Heterogeneous Graphs}

%%
%% The "author" command and its associated commands are used to define
%% the authors and their affiliations.
%% Of note is the shared affiliation of the first two authors, and the
%% "authornote" and "authornotemark" commands
%% used to denote shared contribution to the research.

% \author{Kechi Zhang}
% \authornote{Key Laboratory of High Confidence Software Technologies (Peking University), Ministry of Education, China}
% \authornotemark[3]
% \email{zhangkechi@pku.edu.cn}
% \orcid{0000-0002-3290-0244}
% \affiliation{%
% }

% \author{Wenhan Wang}
% \authornote{Nanyang Technological University, Singapore}
% \authornote{The two authors share equal contribution.}
% \email{wwhjacob@hotmail.com}
% \orcid{0000-0002-0585-2136}
% \affiliation{%
% }

% \author{Huangzhao Zhang}
% \authornotemark[1]
% \email{zhang_hz@pku.edu.cn}
% \orcid{0000-0002-0324-4591}
% \affiliation{%
% }

% \author{Ge Li}
% \authornotemark[1]
% \email{lige@pku.edu.cn}
% \orcid{0000-0002-5828-0186}
% \affiliation{%
% }

% \author{Zhi Jin}
% \authornotemark[1]
% \email{zhijin@pku.edu.cn}
% \orcid{0000-0003-1087-226X}
% \affiliation{%
% }

\author{Kechi Zhang}
\authornote{The two authors share equal contribution.}
\email{zhangkechi@pku.edu.cn}
\orcid{0000-0002-3290-0244}
\affiliation{%
  \institution{Peking University}
  \country{China}
}
\additionalaffiliation{%
  \institution{Key Laboratory of High Confidence Software Technologies (Peking University), Ministry of Education, China}
  \country{China}
}

\author{Wenhan Wang}
\authornotemark[1]
\email{wwhjacob@hotmail.com}
\orcid{0000-0002-0585-2136}
\affiliation{%
  \institution{Nanyang Technological University}
  \country{Singapore}
}

\author{Huangzhao Zhang}
\authornotemark[2]
\email{zhang_hz@pku.edu.cn}
\orcid{0000-0002-0324-4591}
\affiliation{%
  \institution{Peking University}
  \country{China}
}

\author{Ge Li}
\authornotemark[2]
\authornote{Corresponding author.}
\email{lige@pku.edu.cn}
\orcid{0000-0002-5828-0186}
\affiliation{%
  \institution{Peking University}
  \country{China}
}

\author{Zhi Jin}
\authornotemark[2]
\authornotemark[3]
\email{zhijin@pku.edu.cn}
\orcid{0000-0003-1087-226X}
\affiliation{%
  \institution{Peking University}
  \country{China}
}

%%
%% By default, the full list of authors will be used in the page
%% headers. Often, this list is too long, and will overlap
%% other information printed in the page headers. This command allows
%% the author to define a more concise list
%% of authors' names for this purpose.
% \renewcommand{\shortauthors}{Trovato and Tobin, et al.}

%%
%% The abstract is a short summary of the work to be presented in the
%% article.
\begin{abstract}
% When using neural networks to learn representations for program source code, it is beneficial to consider program structure. Most existing code representation learning studies 
% % \zhz{researches or studies; "work" is not countable} 
% use abstract syntax trees (AST) or AST-based graphs to address the programs' structural information. However, these works neglect the node and edge types within ASTs. Different AST nodes convey different types of information, and different AST edges specify different types of relations, so type information vastly exists in all AST nodes and edges. 
% Providing the types of AST nodes and edges can help neural networks better understand the semantics of programs. 
% %Neglecting these semantic types of AST nodes and edges can cause neural networks to confuse semantically different programs.

% In this paper, to jointly address node and edge types in graph representations of code, we leverage the concept of heterogeneous graphs for building graph representations of programs. We propose building heterogeneous program graphs (HPG) from program ASTs and using heterogeneous graph neural networks to learn on HPGs for downstream tasks. We evaluate our approach on two tasks: method name prediction and code classification. Both tasks require reasoning on the semantics of complete code snippets. Experiment results show that our approach outperforms existing sequential or graph-based code representation models, showing that leveraging the type information of nodes and edges in program graphs is helpful for learning program semantics.

Code representation, which transforms programs into vectors with semantics, is essential for source code processing.
We have witnessed the effectiveness of incorporating structural information (\ie, graph) into code representations in recent years.
Specifically, the abstract syntax tree (AST) and the AST-augmented graph of the program contain much structural and semantic information, and most existing studies apply them for code representation.
The graph adopted by existing approaches is homogeneous, \ie, it discards the type information of the edges and the nodes lying within AST. That may cause plausible obstruction to the representation model.
%\me{Most existing code representation learning studies use abstract syntax trees (AST) or AST-based graphs to address the programs' structural information. However, these approaches neglect semantic type information in AST. The type missing issue of the program graph may obstruct the learning procedure of the representation models.}
% However, the existing approaches utilize the homogeneous graph, which does not provide the type information of the nodes nor the edges.
% The type missing issue of the homogeneous graph may obstruct the learning procedure of the representation models.
In this paper, we propose to leverage the type information in the graph for code representation.
To be specific, we propose the heterogeneous program graph (HPG), which provides the types of the nodes and the edges explicitly.
Furthermore, we employ the heterogeneous graph transformer (HGT) architecture to generate representations based on HPG, considering the type of information during processing.
With the additional types in HPG, our approach can capture complex structural information, produce accurate and delicate representations, and finally perform well on certain tasks.
Our in-depth evaluations upon four classic datasets for two typical tasks (\ie, method name prediction and code classification) demonstrate that the heterogeneous types in HPG benefit the representation models. Our proposed HPG+HGT also outperforms the SOTA baselines on the subject tasks and datasets.
% in most cases.

% When using neural networks to learn representations for program source code, it is beneficial to consider program structure. Most previous code representation learning studies use abstract syntax trees (AST) or AST-based graphs to address the programs' structural information. A number of approaches utilize graph neural networks to represent these program graphs. However, these studies adopt homogeneous graph to represent programs, neglecting the node and edge types or using them roughly within ASTs. Ignoring these importance type information in topology will lead to serious semantic confusion between child nodes, such as in binary operators, \etc. This problem obstructs the learning process of existing GNN-based models to a certain degree. 
% % \me{too short}

% In this paper, to jointly address node and edge types in graph representations of code, we leverage the concept of heterogeneous graphs for building graph representations of programs. 
% We propose building heterogeneous program graphs (HPG) according to the abstract syntax description language (ASDL) and using heterogeneous graph neural networks to learn on HPGs for downstream tasks. 
% We evaluate our approach upon four datasets for two tasks: method name prediction and code classification. Experiment results show that our approach outperforms existing sequential or graph-based code representation models, showing that leveraging the type information of nodes and edges in program graphs is helpful for learning program semantics.
\end{abstract}

%%
%% The code below is generated by the tool at http://dl.acm.org/ccs.cfm.
%% Please copy and paste the code instead of the example below.
%%
% \begin{CCSXML}
% <ccs2012>
%  <concept>
%   <concept_id>10010520.10010553.10010562</concept_id>
%   <concept_desc>Computer systems organization~Embedded systems</concept_desc>
%   <concept_significance>500</concept_significance>
%  </concept>
%  <concept>
%   <concept_id>10010520.10010575.10010755</concept_id>
%   <concept_desc>Computer systems organization~Redundancy</concept_desc>
%   <concept_significance>300</concept_significance>
%  </concept>
%  <concept>
%   <concept_id>10010520.10010553.10010554</concept_id>
%   <concept_desc>Computer systems organization~Robotics</concept_desc>
%   <concept_significance>100</concept_significance>
%  </concept>
%  <concept>
%   <concept_id>10003033.10003083.10003095</concept_id>
%   <concept_desc>Networks~Network reliability</concept_desc>
%   <concept_significance>100</concept_significance>
%  </concept>
% </ccs2012>
% \end{CCSXML}

% \ccsdesc[500]{Computer systems organization~Embedded systems}
% \ccsdesc[300]{Computer systems organization~Redundancy}
% \ccsdesc{Computer systems organization~Robotics}
% \ccsdesc[100]{Networks~Network reliability}

%%
%% Keywords. The author(s) should pick words that accurately describe
%% the work being presented. Separate the keywords with commas.

%% A "teaser" image appears between the author and affiliation
%% information and the body of the document, and typically spans the
%% page.
% \begin{teaserfigure}
%   \includegraphics[width=\textwidth]{sampleteaser}
%   \caption{Seattle Mariners at Spring Training, 2010.}
%   \Description{Enjoying the baseball game from the third-base
%   seats. Ichiro Suzuki preparing to bat.}
%   \label{fig:teaser}c
% \end{teaserfigure}

%%
%% This command processes the author and affiliation and title
%% information and builds the first part of the formatted document.
\maketitle

%\TODO{Merge RQ1 and RQ5.}

%\TODO{A/B test.}

% \me{
% To ensure our experiment results are siginificant enough, we conduct A/B-testing between our proposed model and the best performing SOTA model in each dataset. We fully conducted randomized experiments in order to reduce influences other than model itself.
% and all experiment results shows that the \textit{p-value} is lower than the threshold.
% For all results on four datasets, we can be more than 99\% confident that the result is a consequence of changes our model made and not a result of random chance.
% \footnote{The p-values for CSN-Python, Java-small, Python800, Java250 are 0.0000,0.0000,0.0000,0.0036, respectively.
% The obeserved powers are 100\%, 100\%, 100\%, 92.61\%.
% } 
% The testset of each dataset we use is large enough, which makes our results convincing.

% Method Name p=0.0000
% Python800 p=0.0000
% Java250 p=0.0036
% all p < 0.05
% You can be 99\% confident that this result is a consequence of the changes you made and not a result of random chance.
% }

%\TODO{Ablation of reverse edge. Maybe?}

%\TODO{Metrics. Maybe?}

%\TODO{Dataset splition -- following the existing benchmarks.}

%\TODO{Cognac results.}

%\TODO{Acknowledgement.}

% \TODO{Typos checked}
% \TODO{copyright}
% \TODO{author name}

\section{Introduction}
\label{sec:intro}

% \me{enrich motivating examples}
% \me{contributions}
%\zhz{You maybe need to mention something about general code representation, along with the applications, to stress how important this topic is. List some existing work here, including those graph-based or tree-based solutions. After that, talk about the shortcomings of existing work, and how you can tackle them in one or two sentences.}

Generating representations for programs is an essential process in source code processing.
This process transforms programs of different formats (token sequences, abstract syntax trees, or dependency graphs, \etc) into vectorized or tensorized representations. Code representation learning is also the basis of many other tasks, including code summarization \cite{hu2018deep}, method name prediction \cite{alon2018code2seq,DBLP:journals/pacmpl/AlonZLY19}, code classification \cite{DBLP:conf/aaai/MouLZWJ16} and type inference \cite{allamanis2020typilus}, \etc. 
With powerful deep learning (DL) techniques, representations of code can be obtained via a variety of deep neural networks.
In order to obtain better code representations, researchers have incorporated structural information into the modeling process -- from the trivial token sequences \cite{DBLP:journals/neco/HochreiterS97,DBLP:conf/emnlp/ChoMGBBSB14}, to the abstract syntax trees (AST) \cite{DBLP:conf/aaai/MouLZWJ16,DBLP:conf/icse/ZhangWZ0WL19,DBLP:journals/pacmpl/AlonZLY19,alon2018code2seq} and finally to the graphs
\cite{allamanis2018learning,fernandes2018structured,brockschmidt2018generative,yin2018learning,allamanis2020typilus}.
\citea{allamanis2018learning} first propose to leverage graph neural networks (GNN) for learning representation of the programs. Specifically, they create program graphs based on ASTs and use the GGNN model \cite{li2015gated} to learn representations of program graph nodes.
Up to the present, we have witnessed the capability of graph-based techniques, as they could achieve the state-of-the-art (SOTA) performance in bug detection and fixing \cite{dinella2019hoppity}, clone detection \cite{wang2020detecting}, variable misuse prediction \cite{DBLP:journals/pacmpl/WangWGW20}, \etc.

\begin{figure}[t]
 \subfigure[\small AST. The framed part is transformed into (b) and (c)] {
  \label{fig:motivating_p1}     
  \includegraphics[width=0.8\columnwidth]{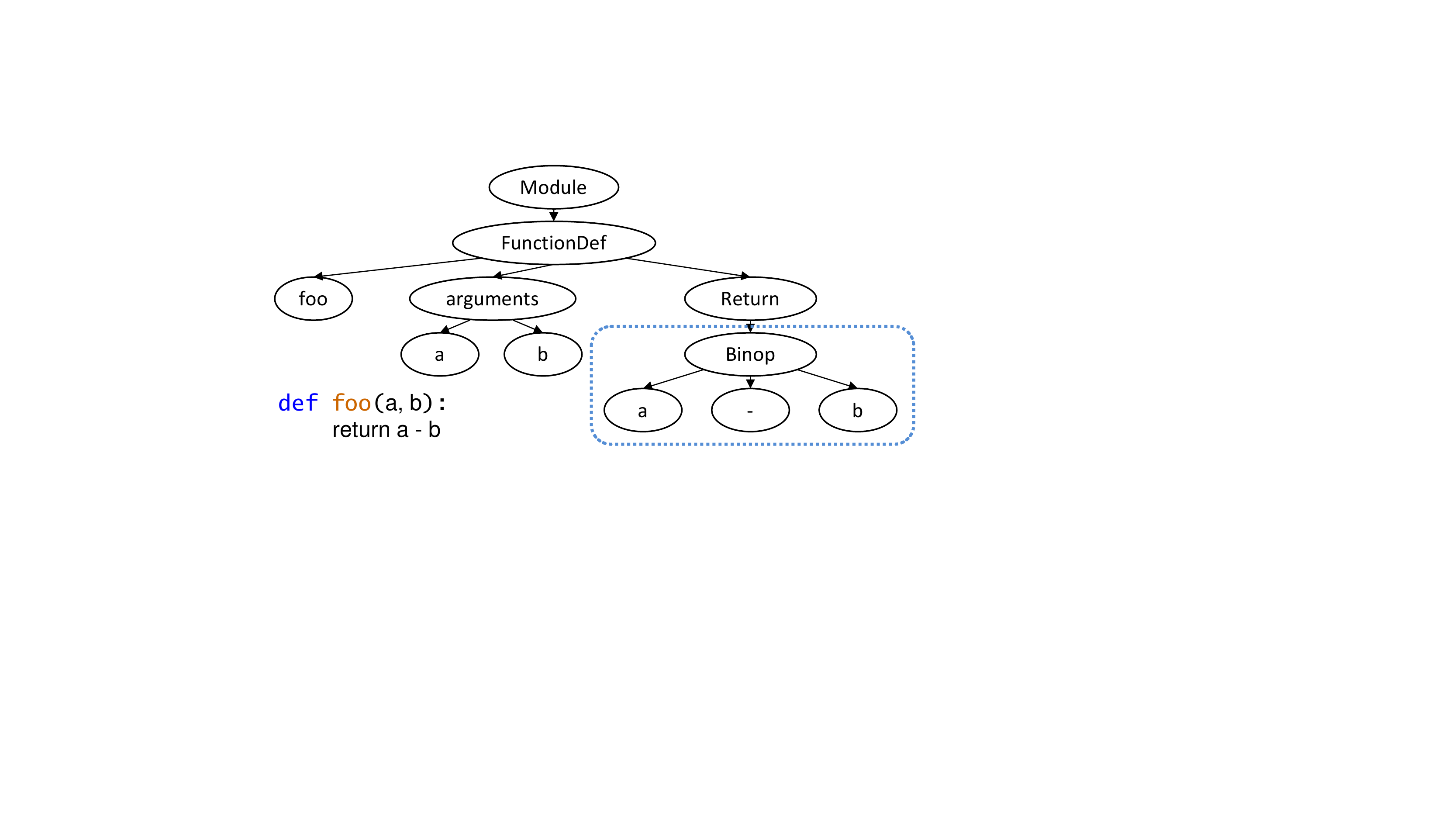}  
  }
 \subfigure[\small Homogeneous graph] {
  \label{fig:binop}     
  \includegraphics[width=0.3\columnwidth]{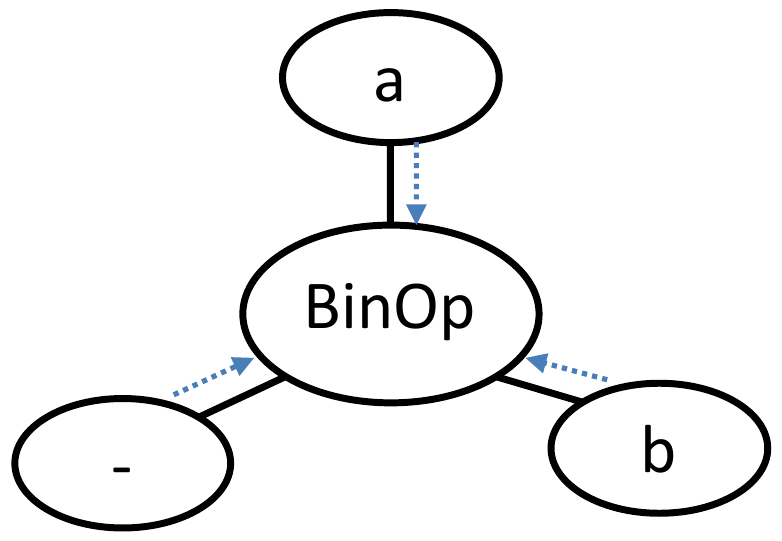}  
  }
   \subfigure[\small Heterogeneous graph] {
  \label{fig:binop_hete}     
  \includegraphics[width=0.57\columnwidth]{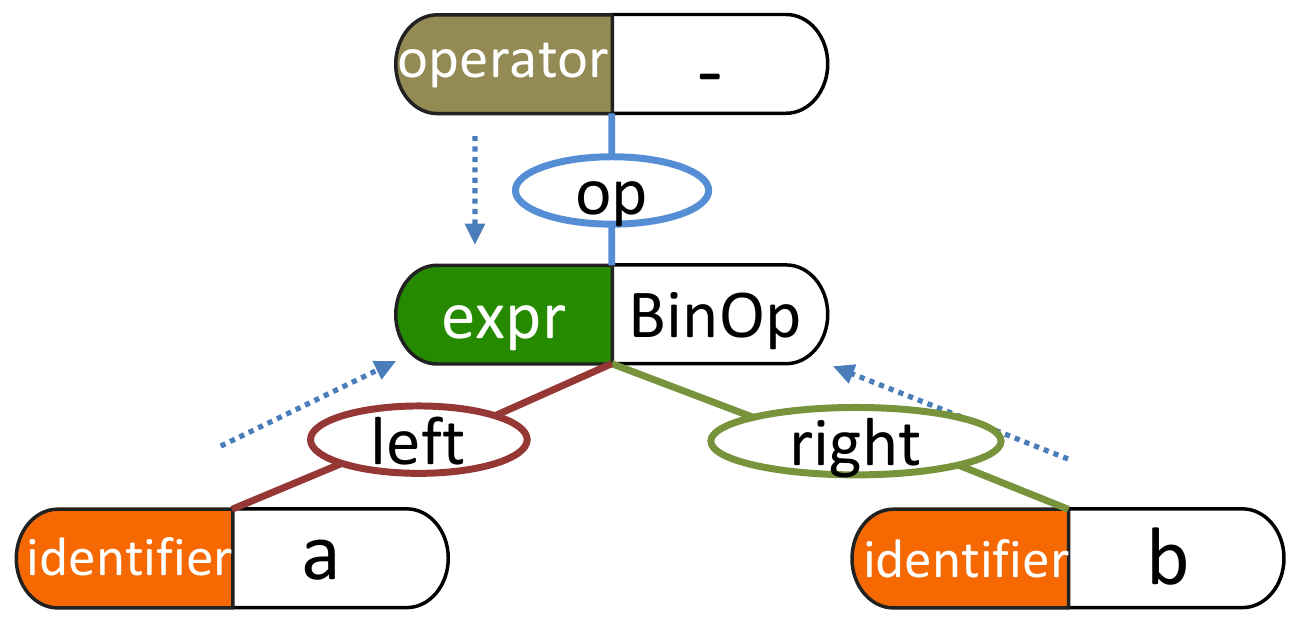}  
  }

\caption{\small An illustrative example of homogeneous and heterogeneous graphs. Previous studies utilize the homogeneous graph as shown in (b) to represent the framed part in (a), while our proposed HPG provides type information of the nodes and the edges as in (c).}
\label{fig:message_passing}
\vspace{-5pt}
\end{figure}

Although the existing graph-based techniques have proven their usefulness, there remains one issue that can not be ignored -- the existing approaches mostly neglect the semantic type information and adopt the (partially) homogeneous graph, \ie, all nodes are of the same type and the edges are only categorized in a coarse-grained manner.
To be clear, previous studies regard the edges in the AST as the same type and augment the AST with other kinds of edges, such as data flow edges \cite{allamanis2018learning}, to retrieve the acquired graph.
The homogeneous graph lacks sufficient type information, and it may make the model incapable of jointly learning the semantics of the nodes and the edges in the graph.
The issue of lacking type information then leads to two major shortcomings in the representation models.
\ding{182} The model cannot explicitly recognize the different type of each node. For instance, in the homogeneous graph in Figure \ref{fig:binop}, the model cannot distinguish the type of {\fontfamily{\ttdefault}\selectfont a}, {\fontfamily{\ttdefault}\selectfont -}, and {\fontfamily{\ttdefault}\selectfont b}, although {\fontfamily{\ttdefault}\selectfont a} along with {\fontfamily{\ttdefault}\selectfont b} are identifiers and {\fontfamily{\ttdefault}\selectfont -} is an operator. Ignoring the node types may hinder the model from sufficiently capturing the semantics of the program.
\ding{183} Topologically, the model regards all connections of nodes (\ie, the edges) to be identical, leaving much structural information discarded. For the same concrete example, when the edge type is not provided, the expressions of ``a-b'' and ``b-a'' may lead to the same homogeneous graph in Figure \ref{fig:binop}. In other words, the semantic meaning of a homogeneous graph can be ambiguous (both ``a-b'' and ``b-a'' are suitable for Figure \ref{fig:binop}).
Both of these two drawbacks are caused by missing information in the homogeneous graph, and it is probable that they obstruct the learning process of the model to a certain degree.

To tackle the issues of the homogeneous graph, in this paper, we propose to leverage heterogeneous graphs for code representation. As the name suggests, the heterogeneous graph is made up of nodes and edges of different types, \ie, there are multiple types of the nodes or the edges, or both. The type information about the nodes and the edges brings forth two advantages, which may settle the aforementioned issues lying within the homogeneous graph.
\ding{182} The types (both node and edge) are explicitly provided in the heterogeneous graph. Heuristically, the semantic information of the nodes and the edges is supposed to be clustered by their types. \Eg, given Figure \ref{fig:binop_hete}, the representation model is informed that {\fontfamily{\ttdefault}\selectfont a} and {\fontfamily{\ttdefault}\selectfont b} are identifier nodes and {\fontfamily{\ttdefault}\selectfont -} is an operator node. Therefore, the model can process these nodes in different manners according to their types.
\ding{183} The edge type guarantees that the graph is not ambiguous. Also take Figure \ref{fig:binop_hete} for instance, the two edges associated with {\fontfamily{\ttdefault}\selectfont a} and {\fontfamily{\ttdefault}\selectfont b} are of type {\fontfamily{\ttdefault}\selectfont left} and {\fontfamily{\ttdefault}\selectfont right} respectively. Hence, Figure \ref{fig:binop_hete} can only refer to the expression of ``a-b'' rather than ``b-a''. This may cause the learning procedure of the model much easier.

In this paper, we propose the heterogeneous program graph (HPG) for code representation. An HPG of a certain code snippet is generated based on its AST according to the abstract syntax description language (ASDL) \cite{wang1997zephyr}, as an example is shown in Figure \ref{fig:binop_hete}. The types in HPG allow the representation model to accurately capture the semantics and the relations of the nodes and the edges. We will explain the technical details of how to produce HPG according to the ASDL grammar in section \ref{sec:HPG}.

To make full use of HPG, we employ the heterogeneous GNN as the representation model, which integrates the type information during processing the graph. In this paper, we adopt the heterogeneous graph transformer (HGT) architecture \cite{hu2020heterogeneous} to generate representations according to our proposed HPG. HGT 
applies the heterogeneous attention to model the edges between the nodes. Especially, during the computation of the attention, HGT uses different weights for different types of the nodes and the edges. Therefore, in short words, HGT is capable to process the nodes and the edges of different types in different manners and to capture the more accurate semantic information.

To demonstrate the usefulness of our proposed HPG and HGT (HPG+HGT), we conduct in-depth evaluations upon four datasets for two tasks (\ie, method name prediction \cite{husain2019codesearchnet, allamanis2016convolutional} and code classification \cite{puri2021codenet}) against currently SOTA baselines, including Code Transformer\cite{DBLP:conf/iclr/ZugnerKCLG21}, Cognac \cite{DBLP:conf/sigsoft/Wang0LM21} and other tree-structured or graph-structured models.  
The experimental results demonstrate that
\ding{182} the heterogeneous types in HPG are capable to improve the performance of the GNN models,
\ding{183} HPG+HGT outperforms current SOTA approaches in most cases,
and \ding{184} the design of each component in HPG+HGT is necessary and may benefit the performance.

The main contributions of this paper are summarized as follow.

\begin{itemize}[leftmargin=*]
    % \item We propose to leverage the heterogeneous graph for code representation. To resolve the type missing issue of homogeneous graph, our proposed HPG explicitly provides the type information of the nodes and the edges.
    % \item We propose to employ the HGT architecture to process our proposed HPG. HGT takes advantage of the heterogeneous types in HPG, and accurately captures the semantics of the nodes and the edges considering their types.
    % \item Our in-depth evaluations on four datasets for two classic code representation tasks certify the usefulness of our proposed HPG+HGT.
    \item To the best of our knowledge, we are the very first to uncover the type missing issue within the homogeneous graph for code representation.
    % \item To the best of our knowledge, we are the first to claim the importance of leveraging AST node and edge types for code representation learning. 
    \item We are the very first to propose leveraging the heterogeneous program graph for code representation. Our proposed HPG is capable to relieve the type missing issue in the homogeneous graph. 
    Specifically, HPG explicitly provides the type information of the nodes and the edges in AST.
    \item We propose a heterogeneous-graph-based framework based on the HGT architecture to process HPG. Our in-depth evaluations on four datasets for two classic code representation tasks certify the effectiveness and the usefulness of our proposed HPG+HGT.
\end{itemize}

\section{Related Work}

\subsection{Code Processing by Deep Learning}

Deep learning has been proved useful in plenty of code processing tasks, which can be roughly divided into two categories, \ie, generation tasks and classification tasks.
Generation tasks take program of different formats (\eg, token sequences, ASTs, data flow graphs, \etc) as input and produce a sequence of information, such as natural language documentations, \ie, code summarization \cite{DBLP:conf/acl/IyerKCZ16,hu2018deep,fernandes2018structured,cai2020tag,ahmad2020transformer}) and method names, \ie, method name prediction \cite{allamanis2016convolutional,alon2018code2seq,DBLP:journals/pacmpl/AlonZLY19,DBLP:conf/icse/Liu0BKKKKT19,DBLP:conf/icse/NguyenPLN20,DBLP:conf/icse/Li0N21, DBLP:conf/kbse/JiangLJ19,DBLP:conf/sigsoft/Wang0LM21,DBLP:conf/iclr/ZugnerKCLG21}, \etc
\citea{allamanis2016convolutional} first propose the convolutional attention networks to predict method names given the body.
\citea{hu2018deep} propose DeepCom to generate comments for methods or functions.
\citea{alon2018code2seq} propose code2seq for method name prediction.
\citea{DBLP:conf/iclr/ZugnerKCLG21} propose Code Transformer to incorporate multiple relations to learning both the structure and the context jointly, improving the performance for multiple generation tasks.
As for classification tasks, the model usually transform the program of different formats into vectorized representations and classify according to the semantics.
Typical classification tasks include code classification \cite{DBLP:conf/aaai/MouLZWJ16,DBLP:conf/aaai/BuiYJ21} and code clone detection \cite{DBLP:conf/icsm/SvajlenkoIKRM14,wang2020detecting}, \etc
\citea{DBLP:conf/aaai/MouLZWJ16} first propose tree-based TBCNN for code functionality classification.
Later, \citea{DBLP:conf/icse/ZhangWZ0WL19} propose ASTNN for code classification and clone detection.
Recently, \citea{hellendoorn2019global} propose GREAT based on the transformer architecture with relational information from code graphs. GREAT has achieved SOTA performance on many classification tasks.

In this paper, to generally demonstrate the capacity of our proposed HPG+HGT, we carry out extensive experiments on both generation and classification tasks, \ie, method name prediction \cite{alon2018code2seq,DBLP:conf/iclr/ZugnerKCLG21} and code classification \cite{puri2021codenet}.

\subsection{GNN for Code Representation}

The graph neural networks (GNN) \cite{gori2005new} is a specialized deep learning model designed to model the graph-structured data, and it is widely employed on many general purposed tasks, such as molecule prediction \cite{gilmer2017neural,DBLP:conf/nips/DuvenaudMABHAA15} and social network analysis \cite{perozzi2014deepwalk,qiu2018deepinf}, \etc
\citea{kipf2016semi} propose GCN by averaging the neighbor of each node to gather information.
\citea{li2015gated} propose GGNN, introducing the gate mechanism from GRU \cite{DBLP:conf/emnlp/ChoMGBBSB14}.

As for the field of code representation, the model need to process the program of different formats and produce a vectorized representation for the certain code processing tasks, as introduced in the last section.
As aforementioned, the program often contain extensive structural information. 
% which the representation model is supposed to accurately capture.
The GNN architecture can naturally extract such structural information from the graph, and hence, a lot of GNN solutions to code representation emerges in the recent years \cite{allamanis2018learning,fernandes2018structured,brockschmidt2018generative,yin2018learning,allamanis2020typilus,si2018learning,wang2020detecting,DBLP:journals/pacmpl/WangWGW20}.
Among them, \citea{allamanis2018learning} first employ GGNN to process program graphs, which is augmented from AST with additional manually crafted edges.
Later, the similar approaches, in which the AST-augmented graph and the GGNN architecture are adopted, are employed to tackle many code processing tasks, including code summarization \cite{fernandes2018structured}, expression generation \cite{brockschmidt2018generative}, code edition \cite{yin2018learning} and type inference \cite{allamanis2020typilus}.
However, the aforementioned approaches ignore the types of the nodes and the edges, resulting in possible ambiguous semantics of the graph representation (as illustrated in section \ref{sec:intro}).

In this paper, to tackle the issue, we introduce the heterogeneous graph with type information of the nodes and the edges for the program graph, which has been demonstrated effective for more general purposed tasks \cite{zhang2019heterogeneous,wang2019heterogeneous,hu2020heterogeneous}.
Specifically, \citea{hu2020heterogeneous} propose the heterogeneous graph transformer (HGT) architecture, utilizing multi-head attention based on meta relations, and it has already achieved SOTA on many web-scale graph tasks.
Inspired by the existing heterogeneous GNN \cite{zhang2019heterogeneous,wang2019heterogeneous,hu2020heterogeneous}, we employ the ASDL grammar to retrieve the types of the nodes and the edges in AST, and build the heterogeneous program graph based on the AST and the additional types.
In this paper, we employ HGT as the backbone of our proposed model.

\section{Preliminary}

To facilitate understanding of HPG and HGT, we present some necessary background knowledge in this section, including the formulation of GNN, heterogeneous graph, and ASDL.

\subsection{Graph Neural Networks}
\label{sec:gnn}

The GNN models have shown promising results for modeling the structural information for program codes \cite{DBLP:journals/tnn/WuPCLZY21,DBLP:journals/corr/abs-2110-09610}.
In this section, we briefly introduce the neural message passing framework \cite{gilmer2017neural,xu2018powerful} of the GNN architecture.
In general, message passing refers to neighborhood aggregation, \ie, each node aggregates features of its neighbors to obtain the new node feature.
Please refer to the formulation as below:

    \begin{align}
a_v^{(k)} &= \mathrm{Aggregate}^{(k)}\left(\{h_u^{(k-1)} : u\in \mathcal{N}(v)   \}\right) \label{eq:gnn_aggregate}\\
h_v^{(k)} &= \mathrm{Combine}^{(k)}\left(h_v^{(k-1)}, a_v^{(k)}\right) \label{eq:gnn_combine}
    \end{align}

\noindent where the superscripts $(k)$ or $(k-1)$ refer to the layers, the subscripts $v$ or $u$ refer to the nodes, $a_v^{(k)}$ and $h_v^{(k)}$ are the aggregated vector and the feature vector of the node $v$ from the $k$-th layer, and $\mathcal{N}(v)$ refers to the neighbors of $v$.
To be specific, the $k$-th layer aggregates (collects) feature of $v$'s neighbors from the $k-1$-th layer, as Eq. \ref{eq:gnn_aggregate} suggests, forming the aggregated vector $a_v^{(k)}$ of $v$ in the $k$-th layer. Then $a_v^{(k)}$ and $h_v^{(k-1)}$ are combined, resulting in the new feature $h_v^{(k)}$ of $v$, as shown in Eq. \ref{eq:gnn_combine}.

Most previous GNN for code representation researches apply GNNs on AST-based or AST-augmented graphs \cite{allamanis2018learning,fernandes2018structured,brockschmidt2018generative,yin2018learning,allamanis2020typilus,wang2020detecting,dinella2019hoppity}, while a few studies explore the non-AST-based graph of code, which are constructed by type dependency \cite{wei2019lambdanet}, code property \cite{zhou2019devign} for specific downstream tasks, \etc.

\subsection{Heterogeneous Graph}
\label{sec:relatedwork_HG}

\paragraph{Heterogeneous graph}
A heterogeneous graph \cite{DBLP:series/synthesis/2012Sun} is a graph consisting of different types of entities, \ie, nodes, and different types of relations, \ie, edges. It is defined as a directed graph $\mathcal{G} = (\mathcal{V}, \mathcal{E}, \mathcal{A}, \mathcal{R})$ where each node $v\in\mathcal{V}$ and each edge $e\in\mathcal{E}$ are associated with their type mapping functions $\tau(v): \mathcal{V} \rightarrow \mathcal{A} $ and $  \phi(e): \mathcal{E} \rightarrow \mathcal{R}$. $\mathcal{A}$ and $\mathcal{R}$ denote the sets of the node type and the edge type, respectively. 
In addition, at least one of the nodes or the edges contains multiple types, \ie $|\mathcal{A}|+|\mathcal{R}|>2$.
Compared with the traditional homogeneous graph, the heterogeneous graph provides additional type information of the nodes and the edges, as Figure \ref{fig:message_passing} demonstrates.

\paragraph{Meta relation}
To facilitate understanding of heterogeneous GNN, we present a brief introduction of meta relation \cite{hu2020heterogeneous}. Meta relation jointly considers the types of the edge along with its source and target nodes.
To be specific, for an edge $e = (s,t)$ connecting the nodes $s$ and $t$, its meta relation is defined as $\langle\tau(s),\phi(e),\tau(t)\rangle$, where $\tau$ and $\phi$ identify the types of $s$ along with $t$, and $e$, respectively.
Meta relation describes the pattern of the connections in the heterogeneous graph, which may help to build more accurate and delicate representations than the node or edge types alone.

\subsection{Abstract Syntax Description Language}
\label{sec:asdl}

\begin{figure}[t]
\subfigure[\small An example of ASDL for Python.]{
\label{fig:asdl_grammar}
\includegraphics[width=0.9\columnwidth]{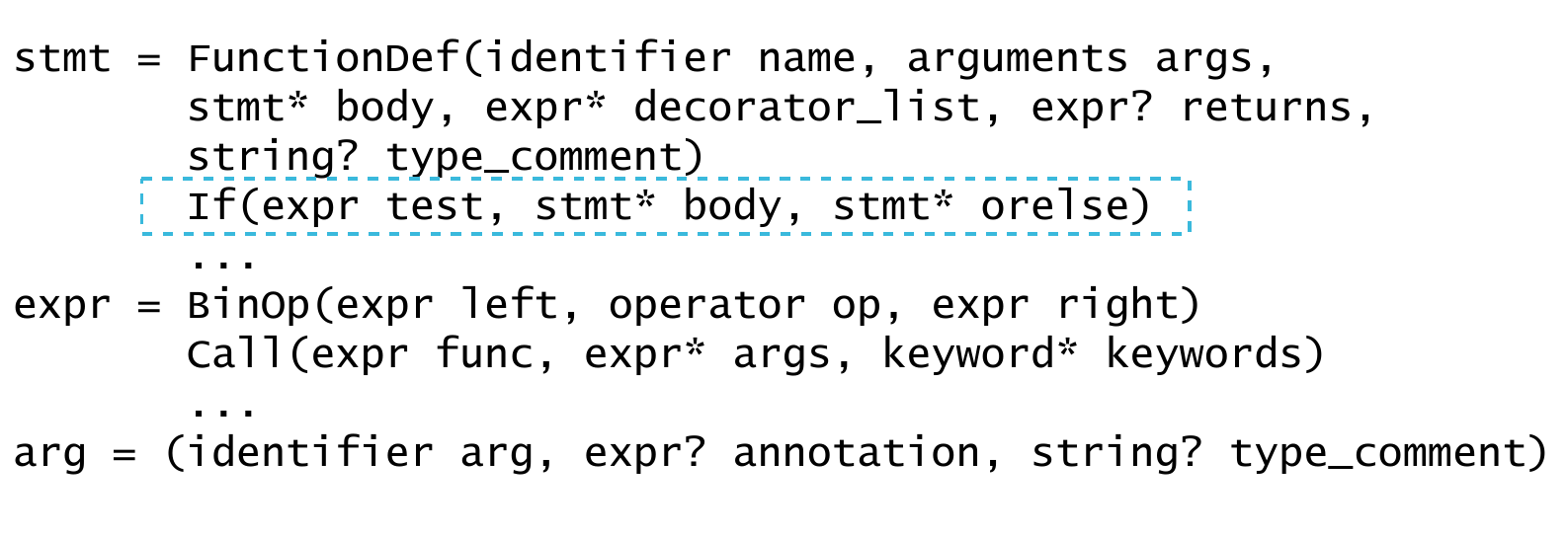}  
}
\subfigure[\small An illustration of a subtree constructed by ASDL.]{
\label{fig:asdl2hpg}
\includegraphics[height=1.0cm]{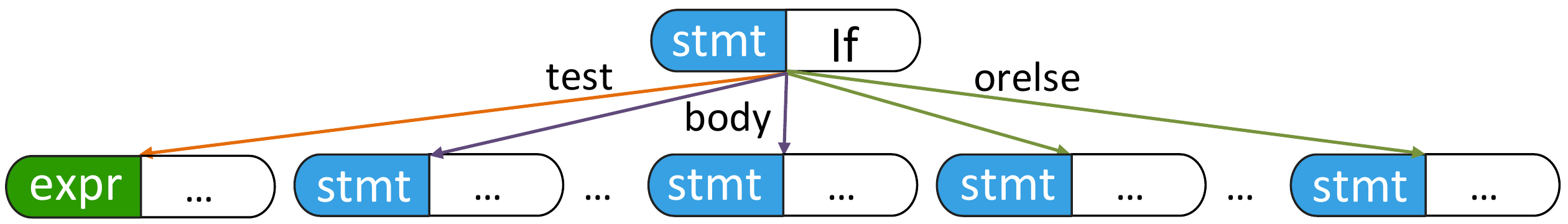}  
}
\caption{\small An illustrative example of ASDL and the syntax tree. The subtree is constructed according to the ASDL rule of {\fontfamily{\ttdefault}\selectfont If}.}
\vspace{-10pt}
\end{figure}

The ASDL grammar contains rich syntactic and semantic information, which has been successfully applied to code generation and semantic parsing \cite{rabinovich2017abstract,yin2018tranx}. 
ASDL \cite{wang1997zephyr}, which consists of a sequence of productions, is quite similar to context-free grammar (CFG). However, ASDL contains additional information, \ie, the node type and the field, which differs it from CFG. The additional information in ASDL can be essential for us to construct heterogeneous graphs for programs. Therefore, we provide a brief introduction to ASDL in this section.

\paragraph{Constructor}
A constructor defines a production rule in ASDL. In general, it defines the parent, the children, and their connections. Take the {\fontfamily{\ttdefault}\selectfont If} constructor in the blue dashed frame in Figure \ref{fig:asdl_grammar} for instance, through it, a subtree in Figure \ref{fig:asdl2hpg} can be produced.

\paragraph{Node type}
As the name suggests, the node type in ASDL is capable to provide the type information of the nodes for the heterogeneous graph.
The node type in ASDL can be divided into two categories -- composite types and primitive types.
\ding{182} The composite type defines a group of constructors, which specifies how to construct such nodes with the certain type. \Eg, in Figure \ref{fig:asdl_grammar}, constructors {\fontfamily{\ttdefault}\selectfont FunctionDef} and {\fontfamily{\ttdefault}\selectfont If}, \etc, are defined by composite type {\fontfamily{\ttdefault}\selectfont stmt}, \ie, the types of these nodes are {\fontfamily{\ttdefault}\selectfont stmt}.
\ding{183} The primitive type, such as {\fontfamily{\ttdefault}\selectfont identifier}, {\fontfamily{\ttdefault}\selectfont int}, and {\fontfamily{\ttdefault}\selectfont string}, \etc, defines a set of terminals.

\paragraph{Field}
The field in ASDL can be viewed as the edge types in the heterogeneous graph.
The field is one of the components in the constructor, and it specifies the relation between the parent and each of its children.
Take the {\fontfamily{\ttdefault}\selectfont If} constructor in Figure \ref{fig:asdl_grammar} for instance, the connection between {\fontfamily{\ttdefault}\selectfont If} and {\fontfamily{\ttdefault}\selectfont expr} is of the field {\fontfamily{\ttdefault}\selectfont test}, and the connenction between {\fontfamily{\ttdefault}\selectfont If} and the first {\fontfamily{\ttdefault}\selectfont stmt} is of the field 
% {\fontfamily{\ttdefault}\selectfont If} between
{\fontfamily{\ttdefault}\selectfont body}. 
According to the field in ASDL, it is handy to mark the edge types in the AST to produce a heterogeneous graph.

\paragraph{Qualifier}
The qualifier in a constructor denotes the number of children in the certain field. There are three valid qualifiers -- single (one and only one), optional ({\fontfamily{\ttdefault}\selectfont ?}, zero or one) and sequential ({\fontfamily{\ttdefault}\selectfont *}, any number). Still take {\fontfamily{\ttdefault}\selectfont If} in Figure \ref{fig:asdl_grammar} for example, it has one {\fontfamily{\ttdefault}\selectfont expr} of the field {\fontfamily{\ttdefault}\selectfont test}, multiple {\fontfamily{\ttdefault}\selectfont stmt}'s of the field {\fontfamily{\ttdefault}\selectfont body}, and multiple {\fontfamily{\ttdefault}\selectfont stmt}'s of the field {\fontfamily{\ttdefault}\selectfont orelse} (see Figure \ref{fig:asdl2hpg}).

\section{Heterogeneous Program Graph}
\label{sec:HPG}

\subsection{Overview of HPG}

\begin{figure}[t]

  \includegraphics[width=0.9\columnwidth]{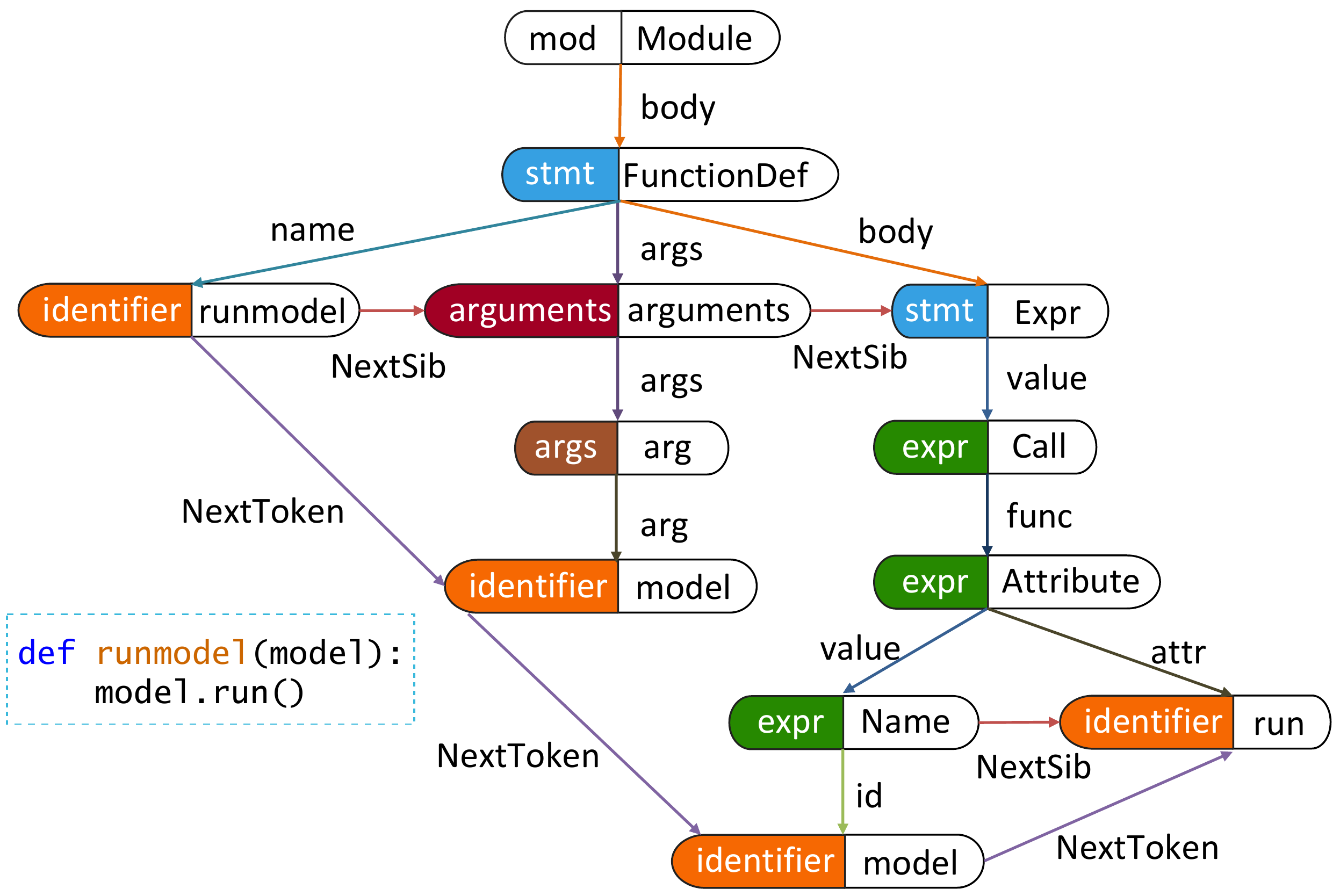} 

\caption{\small An example of HPG for a Python code snippet. Each node is assigned with a type (left part) and a value (right part). Each edge has a type label. We omit the reverse edges for viewing convenience.}
\label{fig:hpg}
\vspace{-10pt}
\end{figure}

The HPG is a format of heterogeneous graph for programs generated from the AST. Formally, the HPG $\mathcal{G}=(\mathcal{V},\mathcal{E},\mathcal{A},\mathcal{R},\mathcal{X})$ consists of a node set $\mathcal{V}$, an edge set $\mathcal{E}$, a node type set $\mathcal{A}$, an edge type set $\mathcal{R}$, and a node value set $\mathcal{X}$. The node $v\in\mathcal{V}$ corresponds to an AST node, which is generated following ASDL. The edges $e_1,e_2,\cdots\in\mathcal{E}$ consists of edges in the ASDL AST and rule-based manually crafted edges. The types of the nodes ($\mathcal{A}$) and the edges ($\mathcal{R}$) are similarly defined as in section \ref{sec:relatedwork_HG}. In addition, for each node $v\in\mathcal{V}$, besides its type $\tau(v)\in\mathcal{A}$, there is also a value $X(v)\in\mathcal{X}$.

Figure \ref{fig:hpg} is an illustrative example of HPG. Each node has a type (left part) and a value (right part), and each edge has a type.
Specifically, the {\fontfamily{\ttdefault}\selectfont NextSib} and {\fontfamily{\ttdefault}\selectfont NextToken} edges are manually crafted edges in addition to the AST edges.

\subsection{Generation of HPG}

With the help of ASDL, we are able to build HPGs from ASTs easily. In general, there are two major steps to generate an HPG -- building a typed AST and inserting manually crafted edges.

\paragraph{Typed AST} To build an AST with its nodes and edges typed, we have to assign the type information according to the ASDL grammar.
\ding{182} For the node, the value is assigned with the corresponding constructor name or the terminal token value,
% which is similar to previous AST-based approaches.
and the type is assigned with the corresponding composite or primitive type (composite for non-terminals and primitive for terminals).
%We take the corresponding composite/primitive type as the node type of non-terminal/terminal nodes.
In addition, for some programming languages (\eg, Python), some nodes only have the type but do not possess a name (\eg, the {\fontfamily{\ttdefault}\selectfont arg} node in Figure \ref{fig:hpg}) \footnote{This phenomenon occurs when a composite type only defines one single constructor. Under such circumstances, the ASDL grammar no longer needs to assign a name to distinguish this certain constructor.}. We set the values of this kind of nodes with their types, \ie, their values and types are identical.
\ding{183} As for the edge, as aforementioned in section \ref{sec:asdl}, the field reflects the relation between the parent and the child, \ie, the edge type. Therefore, we associate each AST edge with their corresponding ASDL field name to build a typed AST.

\paragraph{Manually crafted edge}
In order to retain as much structural information as possible, we also insert heuristic-based manually crafted edges to build HPG, including {\fontfamily{\ttdefault}\selectfont NextSib}, {\fontfamily{\ttdefault}\selectfont NextToken}, and the reversed edges.
\ding{182} {\fontfamily{\ttdefault}\selectfont NextSib} points from a node to its next (right) sibling, and
{\fontfamily{\ttdefault}\selectfont NextToken} points from a terminal node to its next (right) terminal by the order of the text of the code snippet.
We include {\fontfamily{\ttdefault}\selectfont NextSib} and {\fontfamily{\ttdefault}\selectfont NextToken} in HPG because previous work has already demonstrated that these additional edges are quite beneficial to the GNN model \cite{allamanis2018learning,brockschmidt2018generative}. Our experimental results also confirm the necessity of {\fontfamily{\ttdefault}\selectfont NextSib} and {\fontfamily{\ttdefault}\selectfont NextToken}.
\ding{183} The reversed edge $e_{rev}$ of edge $e=(s,t)$, from node $s$ to node $t$, reversely points from $t$ to $s$, \ie, $e_{rev}=(t,s)$. In the HPG, we insert a reverse edge for every forward edge (including the AST edge, {\fontfamily{\ttdefault}\selectfont NextSib}, and {\fontfamily{\ttdefault}\selectfont NextToken}) to improve the connectivity of the graph. The type of the reversed edge is determined by its corresponding forward edge, \eg, the reversed edge of a {\fontfamily{\ttdefault}\selectfont body} edge would be of type {\fontfamily{\ttdefault}\selectfont body\_reverse} and the reversed edge of {\fontfamily{\ttdefault}\selectfont NextToken} would be {\fontfamily{\ttdefault}\selectfont LastToken}. Heuristically, the better the connectivity, the more efficiently and effectively the GNN model can capture the useful features \footnote{During the message passing process, GNN can only consider the information of the $k$-hop reachable neighbors of each node. The one-way forward edge only makes some nodes in the graph unreachable, \eg, no node can reach the AST root in most cases. With the help of the reversed edge, every node is allowed to reach all other nodes. Therefore, during message passing, the $k$-hop neighbor size becomes larger, and the GNN may capture more structural information.}.

\iffalse
\begin{figure}[htbp]

  \subfigure[] {
  \label{fig:a}     
  \includegraphics[width=\columnwidth]{asdlgrammar3.pdf}  
  }
%\framebox[4.0in]{$\;$}
  \subfigure[] {
  \label{fig:b}     
  \includegraphics[width=\columnwidth]{hast2.pdf}  
  }
%\includegraphics[height=8cm, width=12cm]{hast.pdf} 

\caption{An example of the ASDL grammar of Python and an ASDL AST.}
\end{figure}
\fi

\subsection{Integrating Subtoken Information}
\label{sec:subtoken}

\begin{figure}[t]

  \subfigure[\small Shared subtoken scheme] {
  \label{fig:share_subtoken}     
  \includegraphics[height=1.3cm]{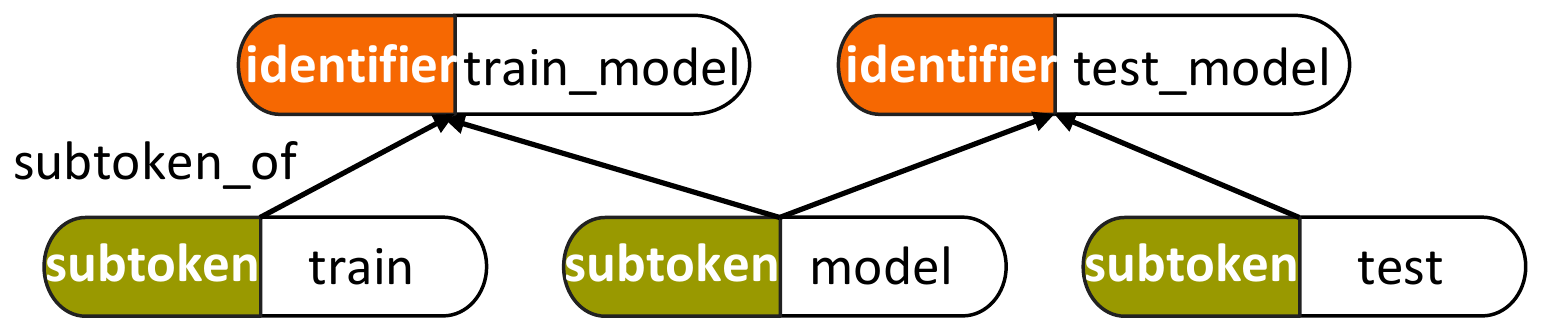}  
  }
%\framebox[4.0in]{$\;$}
  \subfigure[\small Independent subtoken scheme] {
  \label{fig:noshare_subtoken}     
  \includegraphics[height=1.3cm]{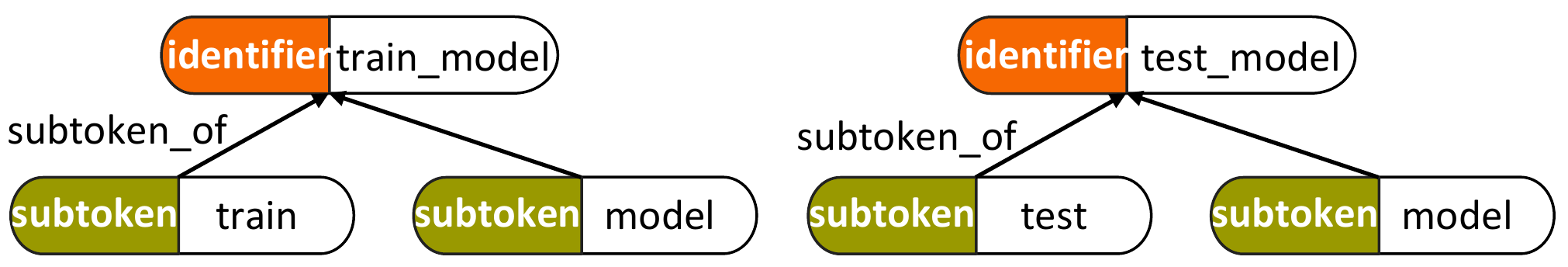}  
  }

\caption{\small Shared and independent schemes for subtoken.}
\label{subtoken_node}
\vspace{-10pt}
\end{figure}

The identifiers, including the method names and the variable names, \etc, are quite essential for code representation because they carry much information in the text as the previous studies suggest \cite{alon2018code2seq,DBLP:conf/icse/NguyenPLN20,DBLP:conf/sigsoft/Wang0LM21}.
The identifier is often composed of several words, increasing the difficulty of extracting useful information. On the other hand, due to the identifiers, the vocabulary size of a program corpus is often extremely large. Therefore, to capture the semantics and to avoid OOV caused by large vocabulary size, we employ the subtoken technique in HPG, by splitting the terminal nodes into multiple parts based on the camel case or the snake case \cite{DBLP:conf/sigsoft/AllamanisBBS15}.

\paragraph{Subtoken}
We introduce a new node type, {\fontfamily{\ttdefault}\selectfont subtoken}, and a new edge type, {\fontfamily{\ttdefault}\selectfont subtoken\_of}, into HPG.
The original terminal nodes are splitted into multiple {\fontfamily{\ttdefault}\selectfont subtoken} nodes by the value, and new {\fontfamily{\ttdefault}\selectfont subtoken\_of} edges are inserted to indicate the origin of the {\fontfamily{\ttdefault}\selectfont subtoken} nodes.
Take Figure \ref{subtoken_node} for illustration, the {\fontfamily{\ttdefault}\selectfont identifier} node with the value ``train\_model'' is splitted into two {\fontfamily{\ttdefault}\selectfont subtoken} nodes -- ``train'' and ``model'', with two {\fontfamily{\ttdefault}\selectfont subtoken\_of} edges inserted from the new {\fontfamily{\ttdefault}\selectfont subtoken} terminal nodes to the original ``train\_model'' node.
In addition, the reversed edge of {\fontfamily{\ttdefault}\selectfont subtoken\_of} is also inserted into HPG during splitting.
Since there may be multiple tokens to be split, we can either share the {\fontfamily{\ttdefault}\selectfont subtoken} nodes among the original {\fontfamily{\ttdefault}\selectfont identifier} nodes, or separate them independently. We will discuss these two schemes in the rest of this section.

\paragraph{Shared subtoken}
In this scheme, the identical {\fontfamily{\ttdefault}\selectfont subtoken} node is shared among different {\fontfamily{\ttdefault}\selectfont identifier} nodes. As demonstrated in Figure \ref{fig:share_subtoken}, the ``model'' {\fontfamily{\ttdefault}\selectfont subtoken} is shared by two {\fontfamily{\ttdefault}\selectfont identifier} nodes, \ie, ``train\_model'' and ``test\_model''. The shared scheme may effectively reduce the size of the graph, and previous work \cite{allamanis2020typilus} has demonstrated that similar strategy is possible to get good semantic representations on the {\fontfamily{\ttdefault}\selectfont subtoken} nodes.

\paragraph{Independent subtoken}
Another scheme is to treat the {\fontfamily{\ttdefault}\selectfont subtoken} nodes of each {\fontfamily{\ttdefault}\selectfont identifier} independently. Take Figure \ref{fig:noshare_subtoken} for illustration, the {\fontfamily{\ttdefault}\selectfont subtoken} nodes of ``train\_model'' and ``test\_model'' are independently separated. The independent scheme may obtain polysemous representations for the same {\fontfamily{\ttdefault}\selectfont subtoken}, but it may suffer from large graph size as there are too many {\fontfamily{\ttdefault}\selectfont subtoken} nodes.

\section{Heterogeneous Graph Transformer}
\label{sec:model_architecture}
% model architecture
% encoder
% decoder
\begin{figure}[t]

  \includegraphics[width=\columnwidth]{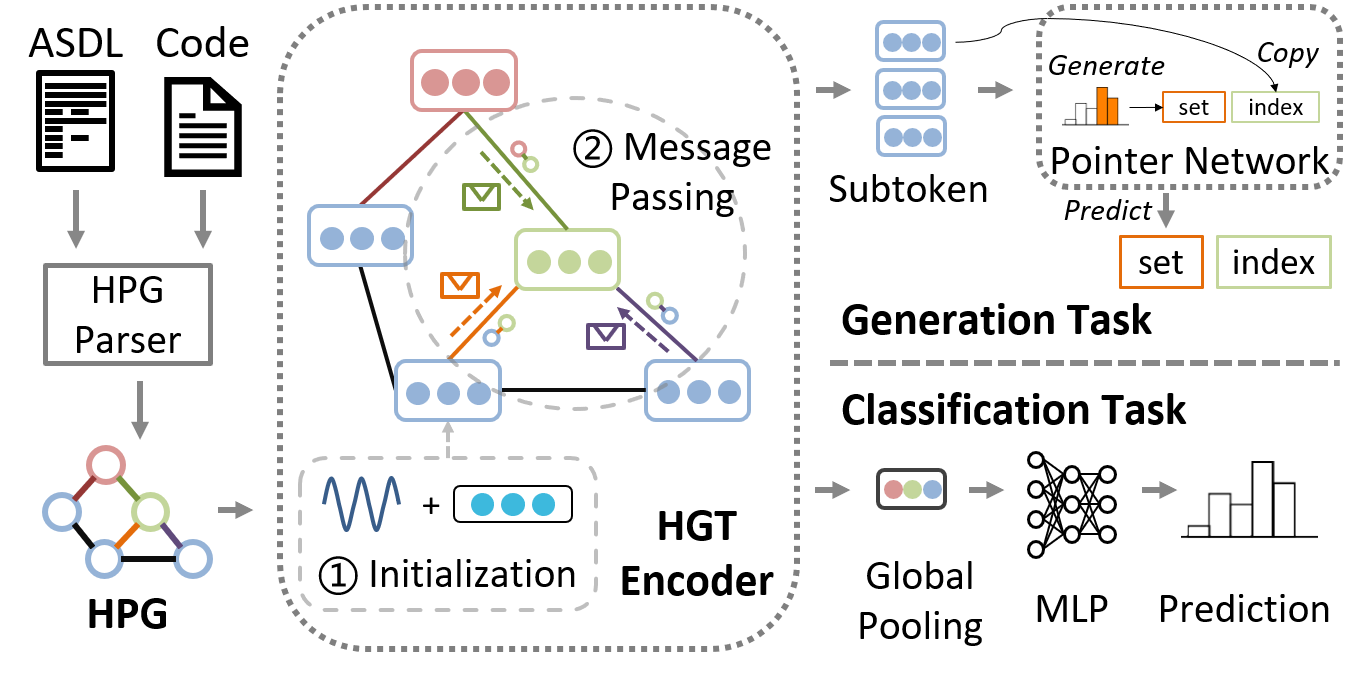}

\caption{\small Overview of the proposed HPG+HGT framework (best view in color). The HPG parser generates HPG from the code snippet according to the ASDL grammar, where the color refers to the types of the nodes and the edges. The HGT encoder takes two steps to process HPG and produce the representations -- feature initialization and iterative message passing (section \ref{sec:hgt_encoder}). At last, the representations are fed into downstream modules for certain tasks (section \ref{sec:hgt4downstream}).
}
\label{fig:model_architecture}
\vspace{-10pt}
\end{figure}

We employ the HGT model, an attention-based heterogeneous graph neural network, as the encoder to generate representations upon the HPG.
HGT encodes HPG to generate a vectorized representation for each node and then performs global pooling to produce the overall representation of the whole HPG.
The representation of the code can be fed into downstream modules for certain tasks.
For generation tasks, such as method name prediction \cite{allamanis2016convolutional,alon2018code2seq,DBLP:journals/pacmpl/AlonZLY19}, we leverage the pointer network \cite{vinyals2015pointer} for sequence generation.
As for classification tasks, such as code classification \cite{DBLP:conf/aaai/MouLZWJ16,puri2021codenet}, we utilize an MLP upon the overall HPG representation as the classifier. 
Figure \ref{fig:model_architecture} presents the overall workflow of our model for different tasks, and we will elaborate on each component in detail in this section.

\subsection{HGT Encoder}
\label{sec:hgt_encoder}

The HGT encoder encodes HPG with the heterogeneous graph transformer architecture \cite{hu2020heterogeneous}.
Specifically, the HGT encoder composes of a positional encoding layer, an embedding layer and multiple HGT layers.
Similar to traditional transformer \cite{vaswani2017attention}, an HGT layer employs heterogeneous mutual attention during message passing, in order to aggregate the information of the previous layer from the neighbors.

\paragraph{Positional encoding}
The technique of positional encoding, which assures the temporal order of the nodes, is essential for transformer-based HGT.
As the program is executed in a fixed order (\eg, different {\fontfamily{\ttdefault}\selectfont stmt} nodes generally have sequential relationships that can not be easily exchanged), modeling order information in HPGs is important for learning their representations. 
We assign a fixed timestamp to each node $v\in\mathcal{V}$, which is the position of $v$ in the depth-first traversal sequence of the AST.
After that, a sine function is conducted upon the timestamps to obtain the positional encoding of each node, which is similar to the previous work \cite{vaswani2017attention,shaw2018self}.

\paragraph{Feature initialization}
It is necessary to have an initialization of the node feature, \ie, $h_v^{(0)}$, to initiate the message passing process. In HGT, we add the embedding of the value and the positional encoding of each node $v$ to obtain $h_v^{(0)}$.

\paragraph{Heterogeneous message}
The heterogeneous message gathers information from the neighbors for each node, considering the types of both the nodes and the edges. Specifically, for a node $t$, we would like to collect information from its neighbor $s\in\mathcal{N}(t)$ ($\mathcal{N}(t)$ is the neighbor node set of $t$), \ie, there exists an edge $e=(s,t)$ pointing from $s$ to $t$. The process is formulated as:

\begin{equation}
    M^{(k)}(s,e,t)=\operatorname{M-Linear}_{\tau(s)}\left(h_s^{(k-1)}\right)\cdot W^M_{\phi(e)} \label{eq:hgt_message}
\end{equation}

\noindent where the information of $s$ ($h_s^{(k-1)}$) is projected into the message space with $\operatorname{M-Linear}_{\tau(s)}$, taking the node type ($\tau(s)$) into account, and then it incorporates the edge type ($\phi(e)$) dependency by $W^M_{\phi(e)}$.
Each $M$ is a message head, and HGT concatenates multiple independent heads forming the heterogeneous message, \ie, $\operatorname{Message}=\operatorname{Concat}(M_1,\cdots,M_h)$ ($(k)$ and $(s,e,t)$ are omitted).

\paragraph{Heterogeneous mutual attention}
The attention determines how important the heterogeneous message is during the aggregation in message passing. Similar to the previous self-attention \cite{vaswani2017attention}, heterogeneous mutual attention computes the attention (importance) score upon the adjacent nodes in HPG, considering the types of the ndoes and the types. To be specific, for $e=(s,t)$, the unnormalized attention score is formulated as:

\begin{equation}
    A^{(k)}(s,e,t)=\left(
            K^{(k)}(s) \cdot
            W^A_{\phi(e)} \cdot
            \left(Q^{(k)}(t)\right)^T
        \right) \cdot
        \frac{\mu_{\langle\tau(s),\phi(e),\tau(t)\rangle}}{\sqrt{d}}
    \label{eq:hgt_attention}
\end{equation}

\noindent where $Q^{(k)}(t)$ and $K^{(k)}(s)$ are the query and the key to compute attention, formulated as $Q^{(k)}(t)=\operatorname{Q-Linear}_{\tau(t)}(h_t^{(k-1)})$ and $K^{(k)}(s)=\operatorname{K-Linear}_{\tau(s)}(h_s^{(k-1)})$, respectively. 
The matrix Multiplication in the parentheses takes the node types ($\operatorname{Q-Linear}_{\tau(t)}$ and $\operatorname{K-Linear}_{\tau(s)}$) and the edge type ($W^A_{\phi(e)}$) into account.
Besides, the trainable prior variable $\mu_{\langle\tau(s),\phi(e),\tau(t)\rangle}$ plays the role of the adaptive scaling factor for each meta relation triplet $\langle\tau(s),\phi(e),\tau(t)\rangle$.
After retrieving the unnormalized score, HGT conduct a softmax activation upon $A$ ($(k)$ and $(s,e,t)$ are omitted, same below) for each $s\in\mathcal{N}(t)$ and get the attention score $\alpha$ of the current attention head.
At last, all independent attention heads are concatenated, forming the heterogeneous mutual attention, \ie, $\operatorname{Attention}=\operatorname{Concat}(\alpha_1,\cdots,\alpha_h)$.

\paragraph{Target-specific aggregation}
After gathering the message and computing the attention score from all neighbors $s\in\mathcal{N}(t)$ of each node $t$, we can thus simply average the messages using the attention scores as the weights. The aggregation is formulated as below:

\begin{equation}
    a^{(k)}_t=\sum_{s\in\mathcal{N}(t)}\left(
            \operatorname{Attention}^{(k)}(s,e,t)
            \cdot
            \operatorname{Message}^{(k)}(s,e,t)
        \right) \label{eq:hgt_combine}
\end{equation}

\paragraph{Residual combination}
Last but not the least, for each node $t$, the aggregated vector $a_t^{(k)}$ is combined with the residual information of $t$ ($h_t^{(k-1)}$), producing the new representation of the current HGT layer. The combination is formulated as:

\begin{equation}
    h_t^{(k)}=\sigma\left(\operatorname{C-Linear}_{\tau(t)}\left(a_t^{(k)}\right)\right)+h_t^{(k-1)}
\end{equation}

\noindent where $\sigma$ is the activation function, and $\operatorname{C-Linear}_{\tau(t)}$ also takes the node type of $t$ ($\tau(t)$) into consideration.

\subsection{Meta Relation in HGT}

\begin{figure}[t]

  \includegraphics[width=0.9\columnwidth]{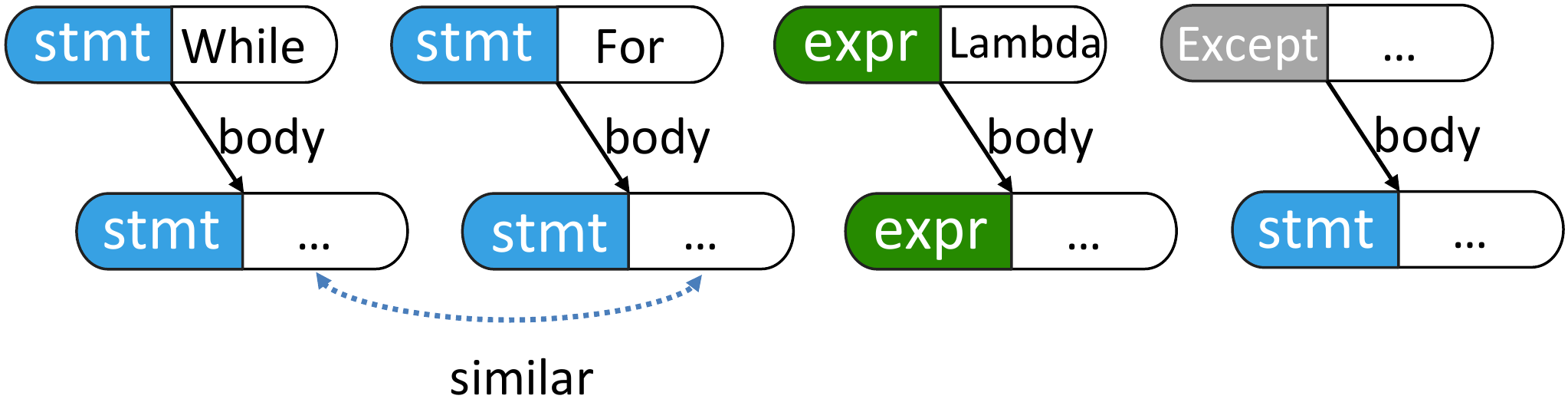}

\caption{\small An illustrative example to demonstrate the effectiveness of the meta relations, which jointly uses the edge types and the node types.
One may easily differ the meta relation of {\fontfamily{\ttdefault}\selectfont stmt}-{\fontfamily{\ttdefault}\selectfont body}-{\fontfamily{\ttdefault}\selectfont stmt} from {\fontfamily{\ttdefault}\selectfont Except}-{\fontfamily{\ttdefault}\selectfont body}-{\fontfamily{\ttdefault}\selectfont stmt}, and therefore the semantics of the {\fontfamily{\ttdefault}\selectfont body} edges pointing to loop bodies and exception handlers can be distinguished.
%Semantic information can be characterized by meta relations like \textit{stmt-body-stmt}, so we can get that the first two examples are more similar.
}
\label{fig:same_edge}
\vspace{-10pt}
\end{figure}

The HGT encoder leverages the types of the nodes and the edges jointly, as illustrated in the last section.
During the message passing process, all $\operatorname{Linear}$ components and all $W$ weights are associated with the types of the nodes or the edges (please refer to the $\tau(s)$, $\tau(t)$ and $\phi(e)$ subscripts in Eq. \ref{eq:hgt_message}-\ref{eq:hgt_combine}).
In addition, the prior $\mu_{\langle\tau(s),\phi(e),\tau(t)\rangle}$ directly models the meta relation triplet.
Therefore, in HGT, different sets of parameters are adopted in the message passing processes for various of meta relations determined by $\langle\tau(s),\phi(e),\tau(t)\rangle$.

Take Figure \ref{fig:same_edge} for illustration, the edges are all with the same type {\fontfamily{\ttdefault}\selectfont body}, but the semantics may differ from each other.
\Eg, the {\fontfamily{\ttdefault}\selectfont body} edges in {\fontfamily{\ttdefault}\selectfont For}-{\fontfamily{\ttdefault}\selectfont body} and {\fontfamily{\ttdefault}\selectfont While}-{\fontfamily{\ttdefault}\selectfont body} both indicates the body of the loop, while {\fontfamily{\ttdefault}\selectfont body} in {\fontfamily{\ttdefault}\selectfont Except}-{\fontfamily{\ttdefault}\selectfont body} refers to an exception handler.
If the types of the nodes and the edges are utilized independently by the model, it is hard to distinguish such differences among the semantics of the {\fontfamily{\ttdefault}\selectfont body} edges.
Fortunately, the meta relation explicitly suggests this kind of semantics differences, as one can easily tells the differences between {\fontfamily{\ttdefault}\selectfont stmt}-{\fontfamily{\ttdefault}\selectfont body}-{\fontfamily{\ttdefault}\selectfont stmt} and {\fontfamily{\ttdefault}\selectfont Except}-{\fontfamily{\ttdefault}\selectfont body}-{\fontfamily{\ttdefault}\selectfont stmt}.
Therefore, HGT is capable of producing accurate and delicate code representations by modeling such meta relations jointly.

\subsection{Downstream Module for HGT}
\label{sec:hgt4downstream}

HGT, serving as an encoder, produces a vectorized representation of the nodes and the whole graph. To complete a variety of downstream tasks, we need to employ the downstream modules.
Most program processing tasks can be categorized into generation tasks and classification tasks.
\ding{182} In generation tasks, such as method name prediction \cite{DBLP:conf/sigsoft/AllamanisBBS15,alon2018code2seq}, the model is supposed to output a sequence of information, \eg, comment texts, method names, \etc, based on the context of the input program (code representation). We equip HGT with the pointer network for generation tasks, which can either generate new words or copy from the context directly.
\ding{183} In classification tasks, such as code classification \cite{DBLP:conf/aaai/MouLZWJ16}, the model makes classification based on the code representation. We connect an MLP classifier to HGT for classification tasks.

\paragraph{HGT for generation tasks}
Generation tasks demand an accurate understanding of the semantics of the input code snippet. Therefore, we follow the classic encoder-decoder architecture \cite{DBLP:conf/emnlp/ChoMGBBSB14}, and utilize a transformer model \cite{vaswani2017attention} to decode based on the representations of the {\fontfamily{\ttdefault}\selectfont subtoken} nodes in HPG from HGT. Besides direct decoding, we also incorporate the pointer network into our model. At each decoding step, the model operates in three steps.
\ding{182} The model collects representations of the {\fontfamily{\ttdefault}\selectfont subtoken} nodes ($s_1,\cdots,s_n$) produced by HGT ($h_{s_1},\cdots,h_{s_n}$), and conducts attention upon them (the attention scores are denoted as $a_1,\cdots,a_n$). In addition, the context vector $h^*$ is also obtained as a weighted average, \ie, $h^*=\sum_{i=1}^na_i\cdot h_{s_i}$.
\ding{183} Based on $h^*$, the model determines the probability of copy from the {\fontfamily{\ttdefault}\selectfont subtoken} nodes ($p_{\operatorname{copy}}$). On the other hand, the probability of generating a new word by the transformer is defined as $p_{\operatorname{gen}}=1-p_{\operatorname{copy}}$.
\ding{184} The probability of decoding the word $w$ is computed as:
    $Prob(w)=p_{\operatorname{gen}}\cdot\operatorname{Transformer}(w)
        +p_{\operatorname{copy}}\cdot\sum_{i:s_i=w}a_i$,
where $\operatorname{Transformer}(w)$ is the probability that the transformer model outputs $w$ and $\sum_{i:s_i=w}a_i$ is the probability that $w$ is copied from the {\fontfamily{\ttdefault}\selectfont subtoken} nodes. $Prob(w)$ is the joint probability of generating and copying $w$ by the model. The copying mechanism empowered by the pointer network alleviates the OOV problem \cite{vinyals2015pointer} of the transformer, by allowing the model to point directly to occurrences of the words. This can effectively improve the performance for those less frequent words during generation.

\paragraph{HGT for classification tasks}
The classification tasks are rather simple, compared with the generation tasks. It involves two steps to complete the classification task with HGT.
\ding{182} We must obtain the overall representation of the whole graph. We apply global attention pooling \cite{li2015gated} over representations of all nodes in HPG.
\ding{183} After receiving the graph representation, we employ an MLP classifier to conduct classification.

\section{Evaluation}

With HPG and HGT, we perform in-depth evaluations upon four datasets against multiple current SOTA baselines to investigate the following research questions:

% \paragraph{RQ1. Heterogeneous Type}
% Is the heterogeneous type information in HPG capable to boost the better representation of the GNN model? To what extent does HPG improve the model performance?

% How effective is the type information in HPGs in improving the performance compared with homogeneous graphs?

\paragraph{RQ1. On Generation Task}
How does HPG+HGT perform on generation tasks?
Specifically, on method name prediction, is HPG+HGT capable to outperform other SOTA approaches?

% how effect does our approach perform compared with the state-of-the-art on method name prediction?

\paragraph{RQ2. On Classification Task}
How does HPG+HGT perform on classification tasks?
Specifically, how does our approach perform compared with the SOTA on the code classification task?

\paragraph{RQ3. Subtoken Scheme}
Can the subtoken technique in HPG improve the performance?
To what extent do the shared scheme and the independent scheme affect the performance of HGT?

% What are the effects of different strategies of integrating subtoken information in HPGs on downstream tasks?

\paragraph{RQ4. Ablation Study}
Is the heterogeneous type information in HPG capable to boost the better representation of the GNN model? To what extent does HPG improve the model performance?
What is the impact of the components in the design of HPG and HGT?
How do they influence the performance of the downstream tasks?

% To what extent do other components in HPGs affect the performance of our approachs on the above two tasks?

\subsection{Experiment Setting}

The experiments are carried out upon one classic generation task (\ie, method name prediction) and one classic classification task (\ie, code classification). For each task, we evaluate upon two widely-studied datasets (four datasets in total). The baseline models for comparison are among the classic models or the SOTA models.

\paragraph{HPG parser}
We implement HPG parsers for two popular programming languages -- Python and Java. The Python parser conforms to the official Python 3.7 ASDL grammar\footnote{https://docs.python.org/3.7/library/ast.html\#abstract-grammar}. Our implemented Python parser can extract up to 23 types of nodes and 71 types of edges (forward). As for Java, we implement an HPG parser based on tree-sitter\footnote{https://tree-sitter.github.io/tree-sitter/}, which parses the code snippet into the parsing tree. We manually define the field type for Java and assign types to the edges in the parsing tree according to our rules.

\begin{table}[t]\small
  \centering
  \caption{\small Statistics of the datasets}
    \begin{tabular}{clccccc}
    \toprule
    % & \multicolumn{1}{l}{CSN-Python} & \multicolumn{1}{l}{Java-small} & \multicolumn{1}{l}{Python800} & \multicolumn{1}{l}{Java250} \\
    
&             & \multicolumn{2}{c}{Method Name Prediction} && \multicolumn{2}{c}{Code Classification} \\
    \cmidrule{3-4} \cmidrule{6-7}
&& CSN-Python & Java-small && Python800 & Java250   \\
    \midrule
\multirow{3}{*}{\rotatebox{90}{Size}} &    Train & 412,178  & 691,974 && 144,000  & 45,000 \\
&    Valid & 23,107  &  23,844 && 48,000  & 15,000 \\
&    Test &  22,176  &  57,088 && 48,000  &  15,000 \\
    \midrule
\multirow{3}{*}{\rotatebox{90}{\# or len}} &    Node & 184.68      & 92.22 && 179.76      & 279.84 \\
&    Edge & 684.21      & 246.77 && 650.14      & 894.56 \\
&    Name  & 2.2
    &  2.5 && - &  - \\
        \midrule
\multirow{2}{*}{\rotatebox{90}{Type}} &    Node   & 17  &  77 && 17  &  77 \\
&    Edge   & 118  & 104 && 116  & 104 \\
    % Avg. identifier nodes   & 5.7  & 49.2  \\
    \bottomrule
    \end{tabular}%
  \label{tab:dataset}
\end{table}

\paragraph{Subject task and dataset}
We evaluate our approach on method name prediction for generation tasks and code classification for classification tasks.
Specifically, for method name prediction, we select CodeSearchNet-Python (CSN-Python) \cite{DBLP:conf/iclr/ZugnerKCLG21} and Java-small \cite{alon2018code2seq,DBLP:conf/iclr/ZugnerKCLG21} datasets, and for code classification, we select Python800 and Java250 \cite{puri2021codenet} datasets. For all these four benchmarks, we utilize the publicly available well-splitted datasets during evaluation.

In method name prediction, a method with its name masked is fed into the model, and the model needs to predict the original method name.
\textbf{CSN-Python} originates from the CodeSearchNet corpus \cite{husain2019codesearchnet}, consisting of around 450K real-world Python methods.
\textbf{Java-small} is a widely-studied benchmark, containing 11 open-source Java projects from GitHub.
We keep the data split of training, validation and testing consistent with existing work \cite{DBLP:conf/iclr/ZugnerKCLG21,alon2018code2seq}.
Those examples that cannot be parsed by our HPG parser are discarded. 
% Table \ref{tab:method_name_dataset} summarizes the statistics of these two datasets.

In code classification, the model needs to predict the category of the given code snippet.
Both \textbf{Python800} and \textbf{Java250} comes from the CodeNet project \cite{puri2021codenet}.
The datasets are from two open judge platforms, and the snippets are categorized by the problem.
The data split is consistent with the original authors\footnote{The dataset split we use in our experiments is the same as stated by the authors which is provided in  https://github.com/IBM/Project\_CodeNet/issues/29}, and we pre-process the samples with our HPG parsers.
% Table \ref{tab:code_classification_dataset} summarizes the statistics of these two datasets.
Table \ref{tab:dataset} summarizes the statistics of these datasets.

\paragraph{Baseline model}
In method name prediction, we compare our proposed HPG+HGT with code2seq \cite{alon2018code2seq}, GREAT \cite{hellendoorn2019global}, XLNet \cite{DBLP:conf/nips/YangDYCSL19},  Code Transformer \cite{DBLP:conf/iclr/ZugnerKCLG21}, Sequence GNN \cite{fernandes2018structured}, Sequence GINN \cite{DBLP:journals/pacmpl/WangWGW20} and Cognac \cite{DBLP:conf/sigsoft/Wang0LM21}. As for code classification, the baselines includes GCN \cite{kipf2016semi}, GIN \cite{xu2018powerful}, GGNN \cite{li2015gated}, Tree-LSTM \cite{DBLP:journals/corr/TaiSM15}, TBCNN \cite{DBLP:conf/aaai/MouLZWJ16}, TreeCaps \cite{DBLP:conf/aaai/BuiYJ21}) and GREAT \cite{hellendoorn2019global}. 
Since CodeNet has already provided simplified parse trees (SPT) for benchmarks, we directly use them for training and evaluation. 
Our selected baseline models contain both classic code representation models (GCN and TBCNN, \etc) and SOTA solutions (GREAT and Code Transformer, \etc).
The required data formats of the baselines cover token sequences (XLNet and Cognac), AST paths (code2seq), ASTs (Tree-LSTM, TBCNN, and TreeCaps), and graphs (GCN, GIN, GGNN, Sequence GINN, Code Transformer, and GREAT).
The architectures of the baselines include RNNs (code2seq and Cognac), convolutional models (TBCNN and TreeCaps), recursive models (Tree-LSTM), GNNs (GCN, GIN and GGNN, Sequence GINN), and transformers (XLNet, Code Transformer, and GREAT).

\paragraph{Performance indicators}
In method name prediction, we list subtoken-level precision, recall, and F1-score over the target sequence (case insensitive) as the indicator, following the previous work \cite{DBLP:journals/pacmpl/AlonZLY19,alon2018code2seq,DBLP:conf/icse/NguyenPLN20}. 
% \TODO{metrics example}
Precision measures the accurate subtoken ratio in the prediction, recall measures the hitting ratio in the target, and F1-score balances precision and recall. \Eg, supposing the target is $\{`train',`graph',`model'\}$ and the prediction is $\{`train',`model'\}$, then the precision is $1.0$, the recall is $\frac23$, and the F1-score is $0.8$
% \zhz{here!}.
%when ${\rm prediction} = (`train',`model')$ and ${\rm target} = (`train',`graph',`model')$, each subtoken in the prediction is correct so the precision is $1.0$, while the prediction hits two subtoken in the target so the recall is $2/3$.
As for code classification, we employ accuracy as the major performance indicator, as CodeNet suggests \cite{puri2021codenet}.

\paragraph{A/B testing}
We also conduct A/B Testing \cite{kohavi2017online} to show the significance of the experimental results. Specifically, we employ A/B testing to decide whether HPG+HGT outperforms the other baselines with confidence scores.

We implement all models with the PyTorch framework along with the DGL library \footnote{https://www.dgl.ai/} and train them on an NVIDIA Tesla V100 16GB GPU.
In our experiment, the HGT encoder consists of 8 HGT layers. We set the embedding size and the hidden size to 256 and 1,024, respectively. We employ 8 heads in each HGT layer. We set the dropout rate to 0.2. We adopt the AdamW \cite{DBLP:conf/iclr/LoshchilovH19} optimizer with the learning rate of $1\times10^{-4}$.
All models are trained from scratch.

\subsection{RQ1: On Generation Task}

\begin{table}[t]\small
  \centering
  \caption{\small Performance of models on CSN-Python (in \%).} 
  \label{tab:csn_python_rst}%
  \begin{threeparttable}
    \begin{tabular}{lcccc}
\hline
\toprule
Model                 & Precision  & Recall     & F1 & p-value\tnote{$\dagger$}         \\
\midrule
GCN                      & 24.77     & 19.97     & 22.12 & $<.0001$    \\
GGNN                     & 24.07     & 19.09     & 21.29  & $<.0001$   \\
% GGNN (w/ edge type)      & 27.31     & 21.94     & 24.33     \\
\midrule
code2seq \cite{alon2018code2seq}                 & 35.79      & 24.85      & 29.34 & $<.0001$     \\
GREAT \cite{hellendoorn2019global}                    & 35.07      & 31.59      & 33.24 & $<.0001$     \\
XLNet \cite{DBLP:conf/nips/YangDYCSL19}                    & 37.39      & 32.01      & 34.49 & $<.0001$     \\
Code Transformer \cite{DBLP:conf/iclr/ZugnerKCLG21}          & 36.41      & \textbf{33.68}      & 34.99 & $<.0001$     \\
% TPTrans                  & 38.45.     & 33.63.     & 35.88      \\
Cognac \cite{DBLP:conf/sigsoft/Wang0LM21}                   & 32.71      & 27.63      & 29.96 & $<.0001$     \\
\midrule
HPG+HGT (no\_subtoken)    & 46.48      & 26.61      & 33.84 & $<.0001$\\
HPG+HGT (independent)         & \textbf{48.44}      & 29.56      & \textbf{36.71} & -     \\
HPG+HGT (shared)           & 47.69      & 28.49      & 35.67 & $<.0001$\\

\bottomrule
\end{tabular}
\begin{tablenotes}
\footnotesize
     \item[$\dagger$] P-values are calculated by comparing with HPG+HGT (independent).
  \end{tablenotes}
  \end{threeparttable}
\end{table}%

\begin{table}[t]\small
  \centering
  \caption{\small Performance of models on Java-small (in \%).}
 \begin{threeparttable}
\begin{tabular}{lcccc}
\hline
\toprule
Model                 & Precision  & Recall     & F1  & p-value\tnote{$\dagger$}       \\
\midrule
Sequence GINN \cite{DBLP:journals/pacmpl/WangWGW20}                & 64.8  & 56.2  & 60.2 & $<.0001$ \\
Sequence GNN \cite{fernandes2018structured}                 & --     & --     & 51.3  & $<.0001$\\
GGNN                           & 40.3  & 35.3  & 36.9 & $<.0001$ \\
\midrule
code2seq \cite{alon2018code2seq}                     & 50.6  & 37.4  & 43.0 & $<.0001$ \\
GREAT \cite{hellendoorn2019global}                        & 53.6  & 46.4  & 49.8 & $<.0001$ \\
XLNet \cite{DBLP:conf/nips/YangDYCSL19}                        & 55.5  & 46.1  & 50.3 & $<.0001$ \\
Code Transformer \cite{DBLP:conf/iclr/ZugnerKCLG21}              & 54.9  & 49.8  & 52.2 & $<.0001$ \\
% \midrule
Original Cognac \cite{DBLP:conf/sigsoft/Wang0LM21} \tnote{$\ddagger$} & 67.1  & 59.7  & 63.2 & --  \\
% \quad \textit{-No prior knowledge}    & -     & -     & 59.3  \\
% \quad \textit{-No callee information} & -     & -     & 57.7  \\
% \quad \textit{-No caller information} & -     & -     & 60.1  \\
% \midrule
Cognac \cite{DBLP:conf/sigsoft/Wang0LM21} %\textit{-No callee information} 
& --     & --     & 57.7 & $<.0001$ \\
% Cognac \cite{DBLP:conf/sigsoft/Wang0LM21}\textit{-No prior knowledge}    & -     & -     & 59.3  \\
\midrule
HPG+HGT (no\_subtoken)        & 62.2  & 56.1  & 59.1 & $<.0001$   \\
HPG+HGT (independent)               & \textbf{65.3}  & \textbf{57.2}  & \textbf{61.0} & -    \\
HPG+HGT (shared)             & 64.7  & 56.1  & 60.1  & $<.0001$  \\

\bottomrule
\end{tabular}
\begin{tablenotes}
\footnotesize
    %  \item Some data of other approaches are extracted from the recent studies \me{need cite}. "-" denotes no relevant information.
%     %  \item Some data of other approaches are extracted from the recent studies \me{need cite}. "-" denotes no relevant information.
     \item[$\dagger$] P-values are calculated by comparing with HPG+HGT (independent).
     \item[$\ddagger$] Cognac incorporates additional Java-specific prior knowledge. We list the original results here, but we do not compare HPG+HGT with it.
  \end{tablenotes}
  \end{threeparttable}
  \label{tab:java_small_rst}%
\end{table}%

To answer RQ1, we evaluate HPG+HGT for generation tasks, compared with Code Transformer, Cognac, along with other baselines, upon CSN-Python and Java-small.
We list the results of HPG+HGT along with the baselines upon CSN-Python and Java-small in Table \ref{tab:csn_python_rst} and \ref{tab:java_small_rst}, respectively.

\paragraph{CSN-Python}
From Table \ref{tab:csn_python_rst}, we find that HPG+HGT outperforms the baseline models on the precision and F1-score indicators. Especially, HPG+HGT with independent subtokens outperforms the current SOTA Code Transformer by 1.72\%, and it outperforms all baselines significantly with high confidence (p-value$<0.0001$).

There are also some exceptions in Table \ref{tab:csn_python_rst}, as the recall indicator of HPG+HGT is lower than the current SOTA GREAT, XLNet, and Code Transformer models.
Our in-depth investigation on the intermediate log reveals that HPG+HGT tends to predict ``short'' method names with only a few subtokens, while the other models prefer to produce ``long'' method names. This may explain why HPG+HGT has a lower recall, as recall reflects the hit rate of the prediction in the ground-truth, regardless of the predicted name length.
This explanation is consistent with the phenomenon where the baseline models have lower precision -- precision measures the ratio of the accurate prediction, and hence the baselines preferring long method names perform poorly on this indicator.

\paragraph{Java-small}
HPG+HGT outperforms the baseline models greatly upon Java-small, as presented in Table \ref{tab:java_small_rst}. Specially, HPG+HGT with independent subtokens even outperforms most baselines by more than 10\% on F1-score (except for GINN and Cognac). The performance improvement is significant with high confidence (p-value$<0.0001$)

In addition, we must elaborate on some technical details of the Cognac \cite{DBLP:conf/sigsoft/Wang0LM21} model. The full Cognac model leverages additional caller-callee information, which is specially designed for the Java programming language alone. However, unlike the heterogeneous types in HPG, for other programming languages, such additional information in Cognac is hardly available. This can explain why Cognac performs poorly upon CSN-Python, which is Python rather than Java.
To demonstrate the general performance of HPG+HGT for generation tasks in general scenarios, we employ the trivial Cognac model as the baseline during evaluation, \ie, the caller-callee information is not provided for both datasets, even though the performance of the original Cognac is also listed in Table \ref{tab:java_small_rst}.
%We must clarify that on Java-small, the full model of Cognac can achieve 63.2\% F1-score \cite{DBLP:conf/sigsoft/Wang0LM21}, which is still the SOTA result.
%To sum up, Cognac leverages language-specific information to achieve the SOTA result of method naming prediction upon Java-small, while our proposed HPG+HGT attempts to produce accurate code representations for generation tasks in general programming languages.

\begin{tcolorbox}[size=title,breakable]
\textbf{Answer to RQ1:} Our proposed approach generally outperforms the current SOTA baseline models on CSN-Python and Java-small method name prediction tasks.
It suggests that HPG+HGT may be capable of producing more accurate and delicate code representations for the generation tasks.
\end{tcolorbox}

\subsection{RQ2: On Classification Task}

% \me{need to edit content}
\begin{table}[t] \small
  \centering
  \caption{\small Performance of models for code classification (in \%).}
\begin{threeparttable}[b]
\begin{tabular}{lccccc}
\toprule
\multirow{2}{*}{Model}              & \multicolumn{2}{c}{Python800} && \multicolumn{2}{c}{Java250} \\
    \cmidrule{2-3} \cmidrule{5-6}
& Acc & p-value\tnote{$\dagger$} && Acc & p-value\tnote{$\dagger$}   \\
\midrule
GCN                   & 91.81 & $<.0001$     && 89.06 & $<.0001$    \\
GIN                   & 93.17 & $<.0001$     && 90.76 & $<.0001$    \\
GGNN                  & 89.92 & $<.0001$    && 88.46  & $<.0001$   \\
\midrule
Tree-LSTM (root) \cite{DBLP:journals/corr/TaiSM15}\tnote{$\ddagger$}     & 93.95 & $<.0001$     && 93.19 & $<.0001$    \\
Tree-LSTM (attention)\tnote{$\ddagger$} & 93.83 & $<.0001$    && 93.71 & 0.0036     \\
TBCNN \cite{DBLP:conf/aaai/MouLZWJ16}                & 91.10 & $<.0001$    && 90.32 & $<.0001$    \\
TreeCaps \cite{DBLP:conf/aaai/BuiYJ21}             & 90.26 & $<.0001$     && 91.42 & $<.0001$    \\
\midrule
GREAT \cite{hellendoorn2019global}                 & 93.30 & $<.0001$    && 93.15 & $<.0001$  \\
\midrule
HPG+HGT (no\_subtoken)    & 93.81 & $<.0001$ && 93.45  & 0.0004     \\
HPG+HGT (independent)       & 94.35 & $<.0001$ && 93.59 & 0.0021 \\
HPG+HGT (shared)         & \textbf{94.99} & -  && \textbf{93.95} & - \\
\bottomrule
\end{tabular}
  \begin{tablenotes}\footnotesize
    %  \item Some data of other approaches are extracted from the recent studies \me{need cite}. "-" denotes no relevant information.
     \item[$\dagger$] P-values are calculated by comparing with HPG+HGT (shared).
     \item[$\ddagger$] Two Tree-LSTM variants: root representation and global attention pooling.
  \end{tablenotes}
  \end{threeparttable}
    \label{tab:codenet_rst}%
\end{table}%
% \zhz{May need to mine some more interesting findings.}

To answer this RQ, we compare the performance of HPG+HGT with the tree-based and graph-based baselines upon the Python800 and Java250 classification tasks.

We present the classification accuracies of HPG+HGT and other baseline models on Python800 and Java250 in Table \ref{tab:codenet_rst}. In most cases, HPG+HGT with the subtoken technique outperforms the baseline models by at least 2\% on both datasets. 
Tree-LSTM performs comparably with HPG+HGT, but HPG+HGT still outperforms it by 1.04\% and 0.24\% on Python800 and Java250, respectively. The p-values (in most cases $<0.0001$) also demonstrate the significant improvement of HPG+HGT compared with the baselines.

We also discover an interesting phenomenon -- the shared subtoken scheme is better for code classification (see Table \ref{tab:codenet_rst}), while the independent scheme is more effective for method name prediction (see Table \ref{tab:csn_python_rst} and \ref{tab:java_small_rst}). We will discuss it in \textbf{RQ3}.

\begin{tcolorbox}[size=title,breakable]
\textbf{Answer to RQ2:} Our proposed approach HPG+HGT outperforms all the
tree-based and graph-based baselines upon Python800 and Java250 datasets. 
It suggests that our proposed heterogeneous-graph-based approach may extract program semantics and handle the classification tasks well.
\end{tcolorbox}

% \me{share > noshare}
\subsection{RQ3: Subtoken Scheme}

To investigate the impact of suchtoken scheme, as we have found earlier, 
we compare the HPG+HGT models with different subtoken schemes, \ie, the shared scheme, the independent scheme, and the no-subtoken scheme, upon all four datasets.

\paragraph{Impact of subtoken}
The {\fontfamily{\ttdefault}\selectfont subtoken} nodes may be beneficial for HPG+HGT in both generation and classification tasks.
In method name prediction, HPG+HGT with subtoken (both the shared scheme and the independent scheme) performs better than HPG+HGT without subtoken, as Table \ref{tab:csn_python_rst} and \ref{tab:java_small_rst} indicates.
Specifically, by adopting the independent subtoken scheme, the F1-score of HPG+HGT increases at least 1.9\% upon both CSN-Python and Java-small datasets.
As for code classification, similar findings can be drawn according to Table \ref{tab:codenet_rst}. Specially, HPG+HGT with shared subtoken scheme outperforms HPG+HGT without subtoken by about 1.2\% on Python800.

% We notice that various downstream tasks can benefit from different strategies of integrating subtoken information.

\paragraph{Subtoken scheme for generation task}
The independent scheme may be more effective for method name prediction than the shared scheme, as in Table \ref{tab:csn_python_rst} and \ref{tab:java_small_rst}, HPG+HGT with the independent scheme outperforms the shared scheme.
There are three plausible reasons for this phenomenon.
\ding{182} The shared scheme may break the order of the {\fontfamily{\ttdefault}\selectfont subtoken} nodes in the original snippet, which can be informative for generation tasks.
\ding{183} For those subtokens frequently appear in the code snippet, the corresponding {\fontfamily{\ttdefault}\selectfont subtoken} nodes occur only once in HPG with the shared subtoken scheme. However, the copy mechanism (the pointer network) in the decoder takes the repeated subtokens into account, and the shared scheme may be unfavorable.
\ding{184} The independent scheme may model the polysemous subtokens, \ie, it allows different representations for the {\fontfamily{\ttdefault}\selectfont subtoken} nodes with the same values. This property is likely to be advantageous to generation tasks.

\paragraph{Subtoken scheme for classification task}
Unlike method name prediction, for code classification, the shared scheme may be favorable than, as presented in Table \ref{tab:codenet_rst}. Specifically, HPG+HGT with the shared scheme slightly outperforms the independent scheme by about 0.7\% and 0.4\% upon Python800 and Java250, respectively.
We assume that the shared {\fontfamily{\ttdefault}\selectfont subtoken} nodes may reflect the association among the identifiers in the snippet, which can boost the performance for classification tasks.

\begin{tcolorbox}[size=title,breakable]
\textbf{Answer to RQ3:} The subtoken technique is capable to improve the performance of HGT for both generation and classification tasks. Different types of tasks may benefit from different subtoken schemes. Specifically, the independent scheme may be good for method name prediction, while the shared scheme may benefit code classification.
\end{tcolorbox}

\subsection{RQ4: Ablation Study}
\label{sec:ablation_study}

\begin{table}[t]\small
  \centering
  \caption{\small Performance of GGNN and HGT (in \%) with and without heterogeneous type information on CSN-Python.}
\begin{tabular}{lcccc}
\toprule
Model   & Precision   & Recall     & F1 &  $\Delta$F1        \\
\midrule
GGNN                     & 24.07     & 19.09     & 21.29 & --     \\
\quad \textit{+ edge type}      & 27.31     & 21.94     & 24.33 & $\uparrow$ 3.04    \\
\midrule
HPG+HGT                              & 48.44       & 29.56      & 36.71  & --    \\
\quad    \textit{- node type}                               & 43.55       & 21.80      & 29.06 & $\downarrow$7.65     \\
\quad    \textit{- edge type}                               & 43.57       & 22.58      & 29.74  & $\downarrow$6.97 \\
\quad    \textit{- node type \& edge type}                 & 43.03       & 21.44      & 28.62 & $\downarrow$8.09 \\
\bottomrule
\end{tabular}
  \label{tab:ablation_hete_type}%
  \vspace{-3mm}
\end{table}%

Ablation studies are carried out to illustrate the role of the components playing in HPG+HGT, including the heterogeneous types, the manually crafted edges in HPG, and the decoding strategies.
Specifically, we include the additional edge types from HPG to verify the performance gaining upon the classic GGNN model, and we remove the types of the nodes and the edges to examine the performance loss of our proposed HPG+HGT.
We also carry out ablations by removing the manually crafted {\fontfamily{\ttdefault}\selectfont NextSib} and {\fontfamily{\ttdefault}\selectfont NextToken} edges, and adopting different decoding strategies.

\paragraph{Impact of heterogeneity}
To demonstrate the impact of heterogeneous types in HPG, we evaluate GGNN and HPG+HGT upon both homogeneous and hetgerogeneous graphs for CSN-Python, as listed in Table \ref{tab:ablation_hete_type}. 
%We list the results of GGNN and HPG+HGT upon homogeneous and heterogeneous graphs for CSN-Python in Table \ref{tab:ablation_hete_type}.
$\Delta$F1 refers to the change of the absolute value of the F1-score indicator compared with the original configuration.
We add additional edge types (from HPG) to GGNN, which originally processes the homogeneous graph \footnote{The GGNN model can leverage the edge type during the message passing process. For each node, GGNN can use the edge-type dependent weight to aggregate the features from its neighbors.}.
Table \ref{tab:ablation_hete_type} indicates that with the help of the additional edge types, the performance of GGNN significantly improves. Especially, the F1-score rises about 3\%. Therefore, GGNN may benefit from the heterogeneous graph.
% \paragraph{GGNN with additional edge type information}
% To demonstrate the usefulness of the heterogeneous type information, we add additional edge types (from HPG) to GGNN, which originally processes the homogeneous graph\footnote{The GGNN model can leverage the edge type during the message passing process. For each node, GGNN can use the edge-type dependent weight to aggregate the features from its neighbors.}.
% As Table \ref{tab:ablation_hete_type} indicates, with the help of the additional edge type, the performance of GGNN significantly improves. Especially, the F1-score rises about 3\%. Therefore, GGNN may benefit from the heterogeneous graph.
On the other hand, we remove the node types or the edge types or both from HPG, to show the performance loss of HGT when the heterogeneous type information is removed from HPG.
We are able to conclude from Table \ref{tab:ablation_hete_type} that either removing the node types or the edge types from HPG would cause the F1-score to drop about 7\%.
% \paragraph{HPG+HGT without node type and edge type}
% On the other hand, we remove the node types or the edge types or both from HPG, to show the performance loss of HGT when the heterogeneous type information is removed from HPG.
% From Table \ref{tab:ablation_hete_type} we can find that removing either the node types or the edge types from HPG would cause the F1-score to drop about 7\%.
Furthermore, when both the edge and the node types are removed, \ie, the graph degenerates to be homogeneous, HGT loses performance greatly, as the F1-score drops 8\%. Therefore, the heterogeneous types in HPG can be quite essential for HGT to generate accurate code representations for tasks such as CSN-Python.

% \begin{tcolorbox}[size=title,breakable]
% \textbf{Answer to RQ1:} For CSN-Python, the classic GGNN gains better performance when the additional edge type information is provided while HPG+HGT loses performance greatly after removing the heterogeneous types.
% Therefore, the heterogeneous type of the nodes and the edges is likely to be beneficial for the graph-based code representation models.
% \end{tcolorbox}

% \subsection{RQ5: Ablation Study}
% \label{sec:ablation_study}

% Ablation studies are carried out to illustrate the role of the components playing in HPG+HGT, including the manually crafted edges in HPG and the decoding strategy over {\fontfamily{\ttdefault}\selectfont subtoken} nodes in HGT.

\begin{figure}[t]
\includegraphics[width=0.65\columnwidth]{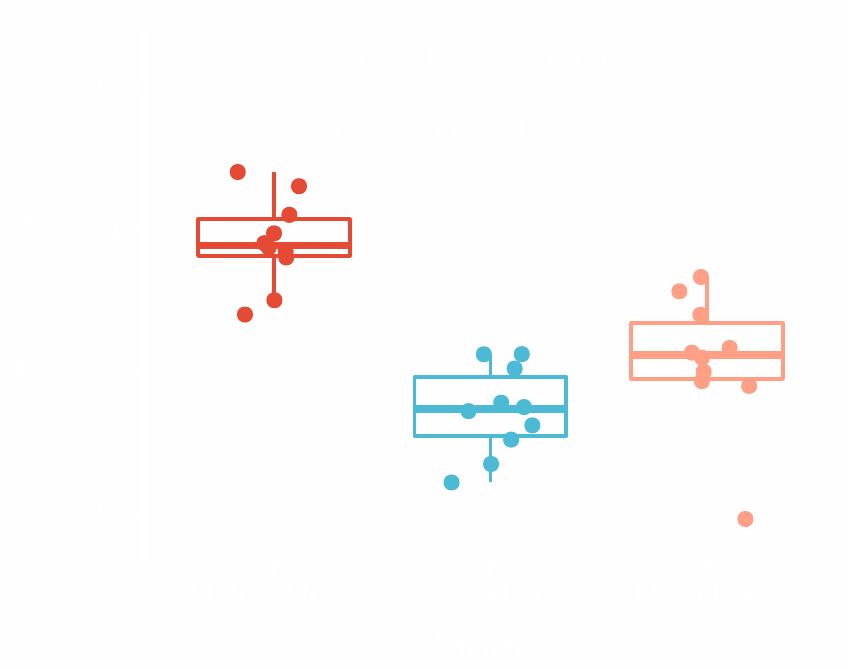}  
\caption{\small Impact of removing the manually crafted {\fontfamily{\ttdefault}\selectfont NextSib} and {\fontfamily{\ttdefault}\selectfont NextToken} edges to HPG+HGT upon Java250 (in \%) with p-values.}
\label{fig:ablation_java250}
\end{figure}

\paragraph{Manually crafted edge}
To demonstrate the usefulness of the manually crafted {\fontfamily{\ttdefault}\selectfont NextSib} and {\fontfamily{\ttdefault}\selectfont NextToken} edges in HPG, we perform an ablation by removing them from HPG.
Figure \ref{fig:ablation_java250} presents the accuracy loss of HPG+HGT with shared subtokens on Java250 when the edges are removed.
% There are two manually crafted edges in HPG, \ie, {\fontfamily{\ttdefault}\selectfont NextSib} and {\fontfamily{\ttdefault}\selectfont NextToken}. Figure \ref{fig:ablation_java250} presents the accuracy loss of HPG+HGT with shared subtokens on Java250 when we remove the manually crafted edges from HPG.
We evaluate the three HGT models for 10 times upon subsets of the testing set.
The accuracy drops about 1.4\% and 1.0\% when {\fontfamily{\ttdefault}\selectfont NextSib} and {\fontfamily{\ttdefault}\selectfont NextToken} are removed respectively.
The p-values ($<0.0002$) indicate that the full HPG+HGT model is significantly better than those without {\fontfamily{\ttdefault}\selectfont NextSib} or {\fontfamily{\ttdefault}\selectfont NextToken} edges upon Java250.
This ablation demonstrates the effectiveness of the manually crafted edges in HPG, and may provide insight into the design of the program graph.

\begin{table}[t]\small
  \centering
  \caption{\small Performance of HPG+HGT with different decoding strategies on CSN-Python (in \%).}
\begin{tabular}{lcccc}
\toprule
Decoding strategy     & Precision   & Recall     & F1 & $\Delta$F1        \\
\midrule
Over {\fontfamily{\ttdefault}\selectfont subtoken} nodes     & 48.44     & 29.56      & 36.71 & --    \\
% \midrule
%     \textit{w/o node type}                               & 43.55       & 21.80      & 29.06     \\
%     \textit{w/o edge type}                               & 43.57       & 22.58      & 29.74 \\
%     \textit{w/o node type and edge type}                 & 43.03       & 21.44      & 28.62 \\
% \midrule
% \quad \textit{w/o NextToken edges}                & 43.07 & 27.18 & 33.33 \\
% \midrule
Over all nodes        & 40.55 & 26.63 & 32.14 & $\downarrow$4.57 \\
\bottomrule
\end{tabular}
  \label{tab:CSN_Python_ablation}%
\end{table}%

\paragraph{Decoding strategy}
As introduced in section \ref{sec:hgt4downstream}, HGT for generation decodes over only the {\fontfamily{\ttdefault}\selectfont subtoken} nodes.
To prove our design, we conduct another ablation upon HPG+HGT with independent subtokens on CSN-Python, where the decoding strategy over all node representations is evaluated.
As table \ref{tab:CSN_Python_ablation} indicates, decoding over all nodes may cause the performance to decrease greatly -- the F1-score decreases about 4.6\% compared with the original design.
To some extent, it verifies our design for generation tasks of the decoding strategy in HPG+HGT.

\begin{tcolorbox}[size=title,breakable]
\textbf{Answer to RQ4:}
Ablation demonstrates the usefulness of the heterogeneous types in HPG.
It also confirms the effectiveness of our design for the manually crafted edges in HPG and the decoding strategy over only the {\fontfamily{\ttdefault}\selectfont subtoken} nodes in HGT.
% Except for heterogeneous types, other components also contribute to its important impact. Specially, manually crafted edges including \textit{NextSib} and \textit{Next} can provide crucial structural information. For method name prediction, decoding over subtoken nodes can significantly boost the performance.
\end{tcolorbox}

\subsection{Threats to Validity}
The dataset selection could be a threat to validity. We counter this by selecting classic method name prediction for generation tasks and code classification for classification tasks. We further select four datasets, which are widely adopted in previous work.
Another threat could be the baseline selection. We counter this by comparing our proposed HPG+HGT with both the classic and the SOTA models.
A further threat could be the language support. We counter this by implementing the HPG parser for the popular Python and Java languages. Our in-depth evaluation, costing more than 3 weeks and over 500 GPU hours, has already demonstrated the effectiveness of HPG+HGT. Since the ASDL grammar is open-sourced, all it takes to build HPG parsers for other programming languages is a large amount of engineering effort. We leave the supporting to more programming languages in future work.

\section{Conclusion}
In this paper, we put forward the idea of heterogeneity in AST and present a framework of representing source code as heterogeneous program graphs (HPG) using the ASDL grammar. By applying heterogeneous graph transformer (HGT) on our proposed HPG, our approach is capable to generate accurate and delicate representations for programs.
Our in-depth evaluations on four classic datasets for two typical code processing tasks (\ie, method name prediction and code classification) demonstrate that the heterogeneous type information in our proposed HPG improves the performance of the representation model.
Furthermore, our proposed HPG+HGT solution even outperforms the SOTA baselines for generation and classification tasks.
%In the future, we plan to evaluate our approach on more tasks, especially node-level or edge-level prediction tasks. We would also extend our approach to other programming languages and propose new models more suited for heterogeneous program graphs.

% \begin{tcolorbox}[size=title]
% \textbf{Answer to RQ1:} blablabl.
% \end{tcolorbox}

\section{Acknowledgement}

% This research is supported by the National Key R\&D Program of China under Grant No. 2020AAA0109400, and the National Natural Science Foundation of China under Grant Nos. 62072007, 61832009, 61620106007. \zhz{Ask Ge Li.}
This research is supported by the National Natural Science Foundation of China under Grant No. 62072007, 62192733, 61832009, 62192730.
We also would like to thank all the anonymous reviewers for constructive comments and suggestions to this paper.

\bibliographystyle{ACM-Reference-Format}
\bibliography{sample-base}

\iffalse
%%
%% If your work has an appendix, this is the place to put it.
\appendix

\section{Research Methods}

\subsection{Part One}

Lorem ipsum dolor sit amet, consectetur adipiscing elit. Morbi
malesuada, quam in pulvinar varius, metus nunc fermentum urna, id
sollicitudin purus odio sit amet enim. Aliquam ullamcorper eu ipsum
vel mollis. Curabitur quis dictum nisl. Phasellus vel semper risus, et
lacinia dolor. Integer ultricies commodo sem nec semper.

\subsection{Part Two}

Etiam commodo feugiat nisl pulvinar pellentesque. Etiam auctor sodales
ligula, non varius nibh pulvinar semper. Suspendisse nec lectus non
ipsum convallis congue hendrerit vitae sapien. Donec at laoreet
eros. Vivamus non purus placerat, scelerisque diam eu, cursus
ante. Etiam aliquam tortor auctor efficitur mattis.

\section{Online Resources}

Nam id fermentum dui. Suspendisse sagittis tortor a nulla mollis, in
pulvinar ex pretium. Sed interdum orci quis metus euismod, et sagittis
enim maximus. Vestibulum gravida massa ut felis suscipit
congue. Quisque mattis elit a risus ultrices commodo venenatis eget
dui. Etiam sagittis eleifend elementum.

Nam interdum magna at lectus dignissim, ac dignissim lorem
rhoncus. Maecenas eu arcu ac neque placerat aliquam. Nunc pulvinar
massa et mattis lacinia.

\fi

\end{document}